\documentclass[]{interact}

\usepackage{epstopdf}% To incorporate .eps illustrations using PDFLaTeX, etc.
\usepackage[caption=false]{subfig}% Support for small, `sub' figures and tables

\usepackage[numbers,sort&compress]{natbib}% Citation support using natbib.sty
\bibpunct[, ]{[}{]}{,}{n}{,}{,}% Citation support using natbib.sty
\renewcommand\bibfont{\fontsize{10}{12}\selectfont}% Bibliography support using natbib.sty
\makeatletter% @ becomes a letter
\def\NAT@def@citea{\def\@citea{\NAT@separator}}% Suppress spaces between citations using natbib.sty
\makeatother% @ becomes a symbol again
\usepackage{multirow}
\usepackage{algorithm}
\usepackage{bm}
\usepackage{url}
\usepackage{algpseudocode}

\theoremstyle{plain}% Theorem-like structures provided by amsthm.sty
\newtheorem{theorem}{Theorem}[section]
\newtheorem{lemma}[theorem]{Lemma}

\theoremstyle{definition}

\theoremstyle{remark}

\begin{document}

\articletype{Research Article}% Specify the article type or omit as appropriate

\title{Bayesian Elastic Net based on Empirical Likelihood}

\author{
\name{Chul Moon\textsuperscript{a}\thanks{CONTACT Chul Moon. Email: chulm@smu.edu} and Adel Bedoui\textsuperscript{b}}
\affil{\textsuperscript{a}Department of Statistical Science, Southern Methodist University, Dallas, Texas, U.S.A.}
\affil{\textsuperscript{b}Department of Statistics, University of Georgia, Athens, Georgia, U.S.A.}
}

\maketitle

\begin{abstract}
We propose a Bayesian elastic net that uses empirical likelihood and develop an efficient tuning of Hamiltonian Monte Carlo for posterior sampling.
The proposed model relaxes the assumptions on the identity of the error distribution, performs well when the variables are highly correlated, and enables more straightforward inference by providing posterior distributions of the regression coefficients.
The Hamiltonian Monte Carlo method implemented in Bayesian empirical likelihood overcomes the challenges that the posterior distribution lacks a closed analytic form and its domain is nonconvex.
We develop the leapfrog parameter tuning algorithm for Bayesian empirical likelihood. We also show that the posterior distributions of the regression coefficients are asymptotically normal. 
Simulation studies and real data analysis demonstrate the advantages of the proposed method in prediction accuracy.
\end{abstract}


	\section{Introduction}

Regularization methods have been introduced to a linear model to improve the prediction accuracy and interpretability by introducing a penalty term to the least squares criterion; lasso that implements $l_1$ penalty \citep{Tibshirani1996}, ridge that adds $l_2$ penalty \citep{Tikhonov1943}, elastic net (EN) that introduces a combination of $l_1$ and $l_2$ penalties \citep{Zou2005}, and non-convex penalties \citep{Fan2001,Zhang2010}. 
It is known that EN shows better performances compared to the lasso while preserving a sparse variable selection property \citep{Zou2005,buhlmann2011statistics}. EN also identifies influential variables better than the lasso and has lower false positives than ridge regression \citep{tutz2009penalized}.

The penalized regression approaches have an intuitive Bayesian counterpart stemming from introducing penalty terms in the form of priors. For instance, Bayesian lasso (BL) \citep{bayeslasso} uses a double-exponential %(or called the Laplace) 
prior, and Bayesian elastic net (BEN) \citep{Ben2010} uses an informative prior that is a compromise between normal and Laplace priors.
The non-convex penalties also have been implemented including spike-and-slab prior \citep{ishwaran2005}, horseshoe prior \citep{Carvalho2010}, and hyperlasso \citep{Griffin2011}.
Motivated by Li and Lin \cite{Ben2010}, various BEN studies have been conducted including applications to gene-expression data  \citep{chen2010detection}, empirical BEN \citep{huang2015empirical}, and regressions \citep{alhamzawi2016bayesian,alshaybawee2017bayesian,alhamzawi2018bayesian}.

The Bayesian approaches on regularization regression have several advantages. First, the Bayesian methods offer a natural interpretation of parameter uncertainty. For example, the standard errors of our parameter of interest can be estimated as the posterior standard deviations because the Bayesian method provides the entire posterior distribution.  
Second, the Bayesian methods provide valid standard errors compared to the variances of the non-Bayesian penalized regressions estimated by the sandwich or bootstrap \citep{kyung2010}.
Third, the penalty parameters in the Bayesian methods can be estimated simultaneously with model parameters. When the non-Bayesian models have multiple penalty parameters, such as EN, multi-stage cross-validations are used to estimate the penalty terms. However, the sequential estimation of penalties may incur the over-shrinkage of parameters that leads to larger bias \citep{Ben2010}. Lastly, the Bayesian penalized regression approaches show similar or better performances compared to the corresponding non-Bayesian approaches \citep{Hans2009,kyung2010,Ben2010}.
%Lastly, the Bayesian penalized regression approaches show similar or better performances compared to the corresponding non-Bayesian approaches \citep{Hans2009,kyung2010,Ben2010}.

Owen \cite{Owen1988,Owen1990} introduces empirical likelihood (EL), a nonparametric likelihood method to make inference for statistical functionals. EL does not require an assumption that data follow a certain distribution while maintaining similar properties of a conventional likelihood method such as Wilk's theorem and Bartlett correctability \citep{Owen1990,Diciccio1991,Chen2006}.
The EL approach has been applied to complex inferences, including general estimating equation \citep{Qin94}, density estimation \citep{Hall1993}, area under the receiver operating characteristic curve \citep{Qin2006}, and data imputation \citep{wang2009}. We refer to Chen and Keilegom \cite{Chen2009} for a review of EL methods on regression.

EL has extended to high dimensional data.
Hjort et al. \cite{hjort2009} and Chen et al. \cite{Chen2009b} show that the asymptotic normality of the EL ratio holds when the dimension of data grows.
Tang and Leng \cite{Tang2010} and Leng and Tang \cite{Leng2012} propose the penalized EL and show that it has the oracle property \citep{Fan2001,fan2004}. Lahiri and Mukhopadhyay \cite{lahiri2012} propose the penalized EL for estimating population means when the dimension of data is larger than the number of observations. Chang et al. \cite{chang2018} also suggest implementing two penalty functions to penalized models and Lagrange multipliers.

Also, the EL-based Bayesian methods have been developed. Lazar \cite{Lazar2003} replaces the likelihood function in Bayesian settings by EL and shows the validity of the resulting posterior distribution.
In addition, Grendar and Judge \cite{Grendar2009} show that Bayesian empirical likelihood (BEL) is consistent under the misspecification of the model. 
BEL has been studied in various areas; the higher-order asymptotic and coverage properties of the posterior distribution for the population mean \citep{Mukerjee2006,Mukerjee2008}, the survey sampling \citep{Wu2010}, small area estimation \citep{Sanjay2011}, quantile regression \citep{Yunwen2012}, Bayesian computation \citep{Mengersen:etal:2012}, sampling method \citep{Chaudhuri2017}, and lasso and ridge regressions \citep{Adel2020}.

    In this paper, we suggest a Bayesian approach for EN where the likelihood function is replaced by the profile EL of the linear model. We place a special prior distribution on the parameter of interest, which combines the $l_1$ and $l_2$ penalties leading to the EN approach. The proposed approach takes advantage of the interpretability and robustness of the results achieved by the Bayesian perspective and EL method, respectively. We implement the HMC sampling for BEL suggested by Chaudhuri et al. \cite{Chaudhuri2017} and propose the leapfrog step size tuning algorithm based on the bisection method.
    The proposed algorithm enables more efficient sampling than hand-tuning or grid-searching HMC parameters. In addition, our method extends the BEL method for penalized regressions of Bedoui and Lazar \cite{Adel2020} by proposing efficient HMC sampling rather than utilizing the tailored Metropolis-Hasting of Chib and Greenberg \cite{Chib1995}.
    
    The outline of the remaining sections is as follows. In Section~\ref{sec:el}, we briefly describe the Bayesian linear model based on EL. In Section~\ref{sec:BNEN}, we propose BEN based on EL and discuss the HMC sampling implementations, along with variable selection procedures. In Section~\ref{sec:sim}, we compare the proposed method with other penalized regression methods using various simulation studies. The air pollution data application is presented in Section~\ref{sec:real}. Section~\ref{sec:conclusion} contains conclusion and future work.
    %``````````````````````````````````````````````````````%
	% Linear Model Based on Empirical Likelihood approach         %
	%``````````````````````````````````````````````````````%
	\section{Linear model based on empirical likelihood}
	\label{sec:el}

We first outline how EL is implemented in a linear model.
Let $X=[\pmb{x_1}, \cdots, \pmb{x_p}]$ be a collection of independent $p$ covariate vectors of size $n$. 
We consider the linear model 
	\[
	y_i=\pmb{x_i}^T\pmb{\theta} + \epsilon_i,
	\]
where the error $\epsilon_i$ is independent and identically distributed and follows an unknown distribution with mean zero and variance $\sigma^2$.
The empirical likelihood function for $\pmb{\theta}$ is defined as
\begin{equation}
L_{EL}(\pmb{\theta})=\sup_{w_i} \Bigl{\{ }  \prod\limits_{i=1}^{n} nw_i \mid w_i \geq 0,\; \sum\limits_{i=1}^{n} w_i=1,\; \sum\limits_{i=1}^{n}w_i\pmb{x_i}(y_i-\pmb{x_i}^{T}\pmb{\theta})=\pmb{0}  \Bigr{\}},
\label{eq4}
\end{equation}
and $\pmb{w}= (w_1,\cdots,w_n)^T$ is the weights on the data points. 
The coefficients $\pmb{\theta}$ can be estimated by maximizing equation (\ref{eq4}) using the Lagrange multipliers. The profile empirical log-likelihood becomes
\begin{equation}
\begin{split}
l_{EL}(\pmb{\theta}) &=-n\log(n) -\sum_{i=1}^{n} \log \left \lbrace  1+\pmb{\gamma}^T\pmb{x_i} (y_i -\pmb{x_i}^T\pmb{\theta})\right \rbrace, \\
& \propto -\sum_{i=1}^{n} \log \left \lbrace  1+\pmb{\gamma}^T\pmb{x_i} (y_i -\pmb{x_i}^T\pmb{\theta})\right \rbrace,\\
\end{split}
\label{eq5}
\end{equation}
where $\pmb{\gamma}=\pmb{\gamma}(\pmb{\theta})$ solves the equation
\begin{equation*}
\sum_{i=1}^{n} \dfrac{\pmb{x_i} (y_i -\pmb{x_i}^T\pmb{\theta})}{1+\pmb{\gamma}^T\pmb{x_i} (y_i -\pmb{x_i}^T\pmb{\theta})} =\pmb{0}.
\label{eqLambda}
\end{equation*}
Here, numerical approaches such as the Newton-Raphson method can be used to find the Lagrange multipliers $\pmb{\gamma}$ \citep{Owen2001}.

The regularization method can be implemented to EL for linear models in different ways. First, the penalty terms can be introduced in the model through the priors $
	\pi(\pmb{\theta}) \propto p_n(\pmb{\theta})
	$, where $p_n$ is a penalty function. Second, Tang and Leng \cite{Tang2010} introduce the penalty terms in EL $l(\pmb{\theta}) 
\propto -\sum_{i=1}^{n} \log \left \lbrace  1+\pmb{\gamma}^T\pmb{x_i} (y_i -\pmb{x_i}^T\pmb{\theta})\right \rbrace - n\sum_{j=1}^{p}p_n(\theta_j)$. In our study, the $l_1$ and $l_2$ penalties are introduced in the form of priors.

	%````````````````````````````````````````````````%
	% Bayesian nonparametric elastic net             %
	%````````````````````````````````````````````````% 
\section{Bayesian elastic net based on empirical likelihood}
\label{sec:BNEN}

 EN uses both the $l_1$ and $l_2$ penalties for $p_t(\pmb{\theta}) $\citep{Zou2005}. The EN estimator $\hat{\pmb{\theta}}_{EN}$ is defined as the minimizer of
	\begin{equation}
	L(\pmb{\theta})= \dfrac{1}{2}||\pmb{y} - \mathit{X}\pmb{\theta}||_2^2 + \lambda_1||\pmb{\theta}||_1 + \lambda_2||\pmb{\theta}||^2_2,
	\label{M2}
	\end{equation}
	
	where
	\[
	\begin{split}
	&\lambda_1, \lambda_2 \geq 0,\\
	& \pmb{\theta}\;\text{is a }p\times 1\;\text{vector},\\
	& ||\pmb{\theta}||_1 = \sum_{j=1}^{p}|\theta_j| \text{ and } ||\pmb{\theta}||^2_2 = \sum_{j=1}^{p}\theta_j^2,\\
	& \pmb{y}\;\text{is a }n\times 1\;\text{vector},\\
	& \mathit{X}\;\text{is a }n\times p\;\text{matrix}.\\
	\end{split}
	\]
	Here, $\lambda_1$ and $\lambda_2$ are $l_1$ and $l_2$ penalty parameters that control the amount of shrinkage. Without loss of generality, we assume that 
		\[
		\sum_{i=1}^{n} x_{ij} =0,\; \sum_{i=1}^{n} y_i=0,\; \sum_{i=1}^{n}x_{ij}^2=1 \text{ for } j=1,\cdots,p.
		\]

    Penalized linear regression models have their Bayesian counterpart models. The Bayesian penalized models introduce priors for $\pmb{\theta}$ where the hyperparameters are functions of the penalty terms. For example, from the form of the EN penalty term in (\ref{M2}), the EN regression parameters can be presented by the following prior
	\begin{equation*}
	\pi(\pmb{\theta}) \propto \exp\left( -\lambda_1||\pmb{\theta}||_1 - \lambda_2||\pmb{\theta}||^2_2\right).
	\label{M3}
	\end{equation*}
Li and Lin \cite{Ben2010} propose the Bayesian counterpart of EN by placing normal and Laplace priors to $l_1$ and $l_2$ penalties, respectively.
	The shrinkage parameters, $\lambda_1$ and $\lambda_2$, are introduced into the model in the form of hyperparameters.

	The BEL approach for linear models replaces the likelihood with EL and places parametric priors. Then, the posterior density of EN under the BEL approach $\pi(\pmb{\theta}|\;\mathit{X},\pmb{y})$ becomes
	\begin{equation*}
	\pi(\pmb{\theta} \mid \mathit{X},\pmb{y})=\dfrac{L_{EL}(\pmb{\theta})\pi(\pmb{\theta})}{\int\limits_{\Theta}^{}L_{EL}(\pmb{\theta})\pi(\pmb{\theta})d\pmb{\theta}} \;\propto\;L_{EL}(\pmb{\theta})\pi(\pmb{\theta}).
	\label{M4}
	\end{equation*}

	The hierarchical representation of the full model becomes
	\begin{equation}
	\begin{split}
	L_{EL}(\pmb{\theta}) & \sim \exp\left( l_{EL} (\pmb{\theta})\right),  \\
    \pmb{\theta}|\sigma^2 &\sim  \exp\left( -\dfrac{1}{2\sigma^2} \left[ \lambda_1||\pmb{\theta}||_1 + \lambda_2||\pmb{\theta}||^2_2\right]\right),\\
	\sigma^2 &\sim  IG(a,b).\\
	\end{split}
	\label{M5}
	\end{equation}
  where $L_{EL}(\pmb{\theta})$ is the profile empirical likelihood for the linear model in (\ref{eq5}), IG is an inverse gamma whose probability density function is $\dfrac{b^a}{\Gamma\left(a\right)} x^{-a -1}\exp\left(-\dfrac{b}{x}\right),$ for $x>0$, and $X$ and $\pmb{y}$ are omitted in $L_{EL}\left(\pmb{\theta}\right)$ for simplicity. 
  We condition $\pmb{\theta}$ on $\sigma^2$ to guarantee the unimodality of the posterior distributions \citep{bayeslasso,Ben2010}. 
	One can also choose a noninformative prior on $\sigma^2$ of the form of $1/\sigma^2$. 
  
  The prior of $\pmb{\theta}$ given $\sigma^2$ defined in (\ref{M5}) is a multiplication of normal and Laplace densities. Andrews and Mallows \cite{andrews1974} show that the Laplace distribution can be represented as a scale mixture of normals with an exponential mixing density
	\begin{equation*}
	\dfrac{a}{2}e^{-a|z|} = \int_{0}^{\infty} \dfrac{1}{\sqrt{2\pi s}} e^{-z^2/(2s)} \dfrac{a^2}{2}e^{-a^2s/2}ds,\;\;\;a\;>\;0.
	\label{eq:laplace2}
	\end{equation*} 
  Li and Lin \cite{Ben2010} also showed that the prior defined in (\ref{M5}) can be presented as a scale mixture of normals with a truncated gamma mixing density. The hierarchical scheme in (\ref{M5}) becomes as follows
	\begin{equation}
	\begin{split}
	L_{EL}(\pmb{\theta}) & \sim \exp\left( l_{EL} (\pmb{\theta})\right),  \\
	\pmb{\theta}|\pmb{\tau},\sigma^2 &\sim  \prod_{j=1}^{p} N\left(0, \left(\dfrac{\lambda_2}{\sigma^2} \dfrac{\tau_j}{\tau_j-1} \right)^{-1}\right),\\
	\pmb{\tau}|\sigma^2 &\sim \prod_{j=1}^{p} TG\left(\dfrac{1}{2}, \dfrac{\lambda_1^2}{8\lambda_2\sigma^2}, (1,\infty)\right),\\
	\sigma^2 &\sim  IG(a,b),\\
	\end{split}
	\label{M6}
	\end{equation}
	where $\pmb{\tau}$ follows the truncated gamma distribution from 1 to $\infty$ with the shape parameter $1/2$ and the rate parameter $\lambda_1^2/\lambda_2\sigma^2$. The full joint posterior density is as follows
	\begin{equation}
	\begin{split}
	\pi\left(\pmb{\theta},\;\pmb{\tau},\,\sigma^2|X,\;\pmb{y}\right) \propto& L_{EL}\left(\pmb{\theta}\right)\times \prod_{j=1}^{p}\left(\dfrac{\lambda_2}{\sigma^2}\dfrac{\tau_j}{\tau_j-1} \right)^{\frac{1}{2}} \exp\left( -\dfrac{1}{2}\dfrac{\lambda_2}{\sigma^2}\dfrac{\tau_j}{\tau_j-1}\theta_j^2\right)\\
	&\times \prod_{j=1}^{p} \left(\frac{\lambda_1^2}{8\lambda_2\sigma^2}\right)^{\frac{1}{2}} \tau_j^{-\frac{1}{2}}\exp\left(-\tau_j\dfrac{\lambda_1^2}{8\lambda_2\sigma^2}  \right)I\left(\tau_j \in (1,\infty) \right) \\
	& \times \left(\frac{1}{\sigma^2}\right)^{a+1} \exp\left(-\frac{b}{\sigma^2}\right).
	\end{split}
	\label{M7}
	\end{equation}
	To sample from the posterior $\pi\left(\pmb{\theta},\;\pmb{\tau},\,\sigma^2|X,\;\pmb{y}\right)$, we draw from the following conditional distributions
	\begin{enumerate}
		\item[1.] Sample $\pmb{\theta}$ from 
		\[L_{EL}\left(\pmb{\theta}\right)\exp\left( -\dfrac{\lambda_2}{2\sigma^2}\sum_{j=1}^{p}\dfrac{\tau_j}{\tau_j-1}\theta_j^2 \right).
		\]
		\item[2.] Sampling $\tau_j$ is equivalent to sampling $\tau_j-1$ from 
		\[
		 \text{GIG}\left(\nu= \dfrac{1}{2},\; \psi=\;\dfrac{\lambda_1^2}{4\lambda_2\sigma^2},\;\chi=\dfrac{\lambda_2}{\sigma^2}\theta_j^2\right) \text{ for } j=1,\cdots,p.
		\] 
		\item[3.] Sample $\sigma^2$ from 
		\[
		\text{IG}\left(a+p,\;b+ \dfrac{1}{2}\sum_{j=1}^{p}\left[ \lambda_2\dfrac{\tau_j}{\tau_j -1}\theta_j^2 + \dfrac{\lambda_1^2}{4\lambda_2}\tau_j \right] \right).
		\]
	\end{enumerate}
	Here, $\text{GIG}(\nu,\;\psi,\;\chi)$ is the generalized inverse Gaussian distribution with probability density function $\dfrac{\left(\psi/\chi\right)^{\nu/2}}{2K_{\nu}\left(\sqrt{\psi\chi} \right)}x^{\nu-1}\exp\left\{-\dfrac{1}{2}\left(\chi x^{-1} +\psi x\right) \right\}$ for $x>0$, where $K_{\nu}\left(\cdot\right)$ is the modified Bessel function of the third kind with order $\nu$. We use the rejection algorithm proposed by H\"{o}rmann and Leydold \cite{Hormann2014} to generate samples from the GIG distribution.

	The posterior estimates of BEN based on EL estimates are consistent.
	As the sample size increases, the estimates converge to the true value of the parameter being estimated. 
	The consistency of the estimators under the BEL framework has been proved for the quantile regression \citep{Yunwen2012} and the penalized regression \citep{Adel2020}, and it can be easily extended to the proposed estimates.
	
	\subsection{Hamiltonian Monte Carlo sampling for Bayesian empirical likelihood}
	
	Implementing traditional sampling methods like Gibbs sampler and Metropolis-Hastings (MH) under the Bayesian empirical likelihood framework is a daunting task. First, the conditional distribution of $\pmb{\theta}$ has a non-closed form, which makes the implementation of the Gibbs sampler impossible. Second, implementing MH is suitable to sample from a distribution that lacks a closed form, but it may fail to achieve convergence. In addition, the nature of the posterior density support complicates the process of finding an efficient proposal density. The support of the posterior EL is nonconvex with many local optima where its surface is rigid. In these cases, the chain can be trapped in a region and not reach the global optimum. Therefore, it is challenging to find the proper proposal density that provides a high acceptance right with the appropriate location and dispersion parameters.

	We use HMC \citep{HMC2011} to sample $\pmb{\theta}$, inspired by Chaudhuri et al. \cite{Chaudhuri2017}.
	Let $\pmb{\theta}$ be the current position vector and $\pmb{m}\sim N(0,M)$ be the momentum vector where $M$ is the dispersion matrix.  
	The Hamiltonian is defined by the joint distribution of $\pmb{\theta}$ and $\pmb{m}$ that can be represented as the the sum of potential energy $U(\pmb{\theta})$ and kinetic energy $K(\pmb{m})$,
	\begin{eqnarray}
	    H(\pmb{\theta},\pmb{m}) &=& U(\pmb{\theta}) + K(\pmb{m}) \nonumber \\
	    &=& - \log  \pi (\pmb{\theta}) + \frac{1}{2}\pmb{m}^TM^{-1}\pmb{m}
	    \label{eq:potential}
	\end{eqnarray}
	The partial derivatives of Hamiltonian determine how the $\pmb{\theta}$ transits to a new state,
	\begin{eqnarray*}
	   \frac{d \pmb{\theta}}{d t} &=& \frac{\partial H}{\partial \pmb{m}} \\
	   \frac{d \pmb{m}}{d t} &=& -\frac{\partial H}{\partial \pmb{\theta}}.
	\end{eqnarray*}
	The Hamiltonian dynamics has reversible, invariant, and volume-preserving properties that enable MCMC updates \citep{HMC2011}. The Hamiltonian equations are computed by discretizing the small time interval $\omega$. The leapfrog integrator is the most widely-used method to implement HMC. 
	First, given the current time $t$ and the position $\pmb{\theta}$, the momentum $\pmb{m}$ is independently drawn from $N(0,M)$. Then the position and the momentum at time $t+\omega$ is updated as follows:
	\begin{eqnarray*}
	 \pi\left( t+\frac{1}{2}\omega \right) &=& \pi(t)-\frac{\omega}{2}\frac{\partial U}{\partial \pmb{\theta}} \\
	 \pmb{\theta}(t+\omega) &=& \pmb{\theta}(t) + \omega M^{-1}\pi\left( t+\frac{1}{2}\omega \right) \\
	 \pi\left( t+\omega \right) &=& \pi\left( t+\frac{1}{2}\omega \right) -\frac{1}{2}\frac{\partial U}{\partial \pmb{\theta}}(\pmb{\theta}(t+\omega)).
	\end{eqnarray*}
	The proposed state $(\pmb{\theta^*},\pmb{m^*})$ is obtained after repeating the above updates $T$ times. Here, $\omega$ and $T$ are also called the step size and leapfrog steps, respectively. The proposed state is accepted with the probability
	\begin{equation*}
	    \min \left[ 1, \exp (-H(\pmb{\theta^*},\pmb{m^*})+H(\pmb{\theta},\pmb{m})) \right].
	\end{equation*}
	
	HMC is known to be more efficient than random-walk-based MH in sampling from the posterior for Bayesian EL. First, Chaudhuri et al. \cite{Chaudhuri2017} show that once the parameters are inside the support, the HMC chain does not go outside and return to the interior of the support if they reach its boundary under certain assumptions. This is due to the property of $l_{EL}(\pmb{\theta})$, whose gradient diverges at the boundary. Second, HMC converges quickly towards the target distribution and enables faster convergence. In HMC, distances between successively generated points are large. Thus, fewer iterations are required to obtain a representative sample.

	The performance of an HMC depends on its parameters, and it is known that tuning them is important \citep{HMC2011, hoffman2014no}. However, the optimal HMC parameter tuning procedure for BEL is not discussed in Chaudhuri et al. \cite{Chaudhuri2017}. It is generally suggested to set a sufficiently small leapfrog step size $\omega$ to ensure that the discretization is good and use sufficiently large leapfrog steps $T$ so that the overall trajectory length $\omega T$ is not too short. However, setting a large $T$ could be inefficient for HMC used in the BEL framework. This is because each leapfrog step requires computationally expensive EL computation. Therefore, we fix the leapfrog steps to $T=10$ and find the optimal step size $\omega$ in our study. 
	
	We develop the leapfrog step size tuning algorithm for BEL based on the bisection method \citep{burden2015numerical}. The optimal step size will achieve the known theoretical optimal acceptance rate of 65.1\% \citep{beskos2013optimal}. For a fixed $T$, a larger step size tends to lower an acceptance rate and vice versa. For given lower and upper tolerance levels for acceptance rates, $\iota_l$ and $\iota_u$, respectively, the proposed algorithm searches for $\omega$ that results in a high acceptance rate, greater than $0.651+\iota_u$, while increasing it by $\varepsilon$. If the acceptance rate is within the target range $[0.651-\iota_l,0.651+\iota_u]$, then $\omega$ is selected. On the other hand, if the step size makes the acceptance rate $>0.651+\iota_u$, the method bisects the interval $[\omega-\varepsilon,\omega]$. The bisection procedure is repeated until the acceptance rate reaches the target range. Compared to the grid search algorithm, the proposed algorithm converges to the target acceptance rate range linearly and enables finer and computationally efficient step size tuning. The detailed bisection algorithm is given in Algorithm~\ref{alg:cap}. One can also consider implementing the No-U-Turn sampler (NUTS) \citep{hoffman2014no} that automatically selects the leapfrog steps and step size. We discuss this in Section~\ref{sec:conclusion}.

	\begin{algorithm}
\caption{Bisection leapfrog step size tuning algorithm for Bayesian EL}\label{alg:cap}
\begin{algorithmic}
\Require itermax, (initial) $\omega$, $T$, and lower and upper tolerances $\iota_l$ and $\iota_u$ 
\Ensure acceptance rate and (updated) $\omega$
\State $\varepsilon \gets \omega$
\State scounter $\gets 0$
\State iter $\gets 1$

\While{$\text{iter} \leq \text{itermax} $}
\State Simulate HMC chain using step size $\omega$ for a given $T$ and compute acceptance rate
\If{acceptance rate $ \in  [0.651 - \iota_l, 0.651 + \iota_u]$}
    \State \textbf{break}
\ElsIf{acceptance rate $ >  (0.651 + \iota_u)$}
    \If{scounter = 0}
    \State $\varepsilon \gets \varepsilon$ \Comment{Do not change $\varepsilon$ if  $\omega$ has not decreased before}
    \Else 
    \State $\varepsilon \gets \varepsilon/2$ 
    \EndIf
\State $\omega \gets \omega + \varepsilon$ \Comment{Update $\omega$ by increasing $\varepsilon$}
\ElsIf{acceptance rate $ <  (0.651 - \iota_l)$}
    \State scounter $+= 1$
    \State $\varepsilon \gets \varepsilon/2$
    \State $\omega \gets \omega - \varepsilon$ \Comment{Update $\omega$ by decreasing $\varepsilon$}

\EndIf
\State iter $+=1$
\EndWhile
\end{algorithmic}
\end{algorithm}

	We run a single chain of length 2,000 with 1,000 burn-ins for quicker step size tuning. For the estimation of $\pmb{\theta}$, we simulate four chains of length 2,000 with 1,000 burn-ins, respectively. 
	
	The convergence of simulated chains are examined with the split-$\widehat{R}$ of Vehtari et al. \cite{vehtari2021rank}
	\begin{eqnarray*}
	    \widehat{R}=\sqrt{\frac{\widehat{\text{var}}^{+}(\pmb{\theta}\mid X,\pmb{y})}{W}},
	\end{eqnarray*}
	where $\widehat{\text{var}}^{+}(\pmb{\theta}\mid X,\pmb{y})=\frac{N_s-1}{N_s} W + \frac{1}{N_s} B$, $W$ and $B$ are the within- and between-chain variances, and $N_s$ is the number of draws of one chain. Followed by Vehtari et al. \cite{vehtari2021rank}, we use the posterior samples that satisfy the convergence criteria $\widehat{R}<1.01$.

	%````````````````````````````````````````````````%
	% Choosing the Bayesian elastic net parameters   %
	%````````````````````````````````````````````````%
\subsection{Choosing Bayesian elastic net penalty parameters}
    \label{subsec:pp}
    The proposed EL based approach needs to specify the penalty parameters $\pmb{\lambda}=(\lambda_1,\lambda_2)$. We consider two approaches: empirical Bayes (EB) and full Bayes (FB). 
    
    First, the EB method estimates the penalty parameter from data and plugs the estimated parameters into the model.
    Park and Casella \cite{bayeslasso} use the EB method to estimate the shrinkage parameter in BL. 
    It treats the parameters as missing data and uses the Monte Carlo expectation-maximization (EM) algorithm approach proposed by Casella \cite{Casella2001}. 
    The EM algorithm is iteratively used to approximate the parameters of interest by substituting Monte Carlo estimates for any expected values that cannot be computed explicitly.
	For our proposed model, $\pmb{\theta},\;\pmb{\tau},$ and $\sigma^2$ are treated as missing data whereas $\lambda_1$ and $\lambda_2$ are treated as fixed parameters.
	
	We use a building block of HMC and the Gibbs sampler to sample $\pmb{\theta}$, $\pmb{\tau}$ and $\sigma^2$. The negative log empirical posterior density of $\pmb{\theta}$ in (\ref{eq:potential}) is defined as
		\[
	-\log\left\{ \pi\left(\pmb{\theta}|\sigma^2,\tau_1^2,\cdots,\tau_p^2 \right)\right\} =\sum_{i=1}^{n}\log \left\{ 1+\pmb{\gamma}^T\pmb{x_i}\left( y_i-\pmb{x_i}^{T}\pmb{\theta} \right) \right\}+\dfrac{\lambda_2}{2\sigma^2}\pmb{\theta}^T\mathit{D}_{\tau}^{-1}\pmb{\theta},
	\]
	where $\mathit{D}_{\tau} = \text{diag}\left( \dfrac{\tau_1}{\tau_1 -1},\cdots, \dfrac{\tau_p}{\tau_p -1}\right)$ and its gradient is defined as
\[
	-\dfrac{\partial \log\left\{ \pi\left(\pmb{\theta}|\sigma^2,\tau_1^2,\cdots,\tau_p^2 \right)\right\}}{\partial \pmb{\theta}}=\sum_{i=1}^{n} \dfrac{-\pmb{\lambda}^T\pmb{x_ix_i}^T}{1+\pmb{\gamma}^T\pmb{x_i}\left( y_i-\pmb{x_i}^{T}\pmb{\theta}\right) }+ \dfrac{\lambda_2}{\sigma^2}\pmb{\theta}^T\mathit{D}_{\tau}^{-1}.
	\]

    The hierarchical model presented in (\ref{M7}) yields the complete-data log-likelihood
    \begin{equation*}
        p\log(\lambda_1)-\frac{\lambda_2}{2\sigma^2}\sum\limits_{j=1}^{p} \frac{\tau_j}{\tau_j-1}\theta_j^2 - \frac{\lambda_1^2}{8\lambda_2\sigma^2} \sum\limits_{j=1}^p \tau_j+\text{terms not involving $\lambda_1$ and $\lambda_2$}.
    \end{equation*}
    At the $k$th step of the Monte Carlo EM algorithm, the conditional log-likelihood on $\pmb{\lambda}^{(k-1)}=\left( \lambda_1^{(k-1)},\lambda_2^{(k-1)} \right)$ and $\pmb{y}$ is
    \begin{equation*}
    \begin{split}
        &Q(\pmb{\lambda}|\pmb{\lambda}^{(k-1)})\\
        &= p\log(\lambda_1)-\frac{\lambda_2}{2}\sum\limits_{j=1}^{p} E\left[ \frac{\tau_j}{\tau_j-1} \frac{\theta_j^2}{\sigma^2} \middle| \pmb{\lambda}^{(k-1)},\pmb{y} \right] - \frac{\lambda_1^2}{8\lambda_2} \sum\limits_{j=1}^p  E\left[\frac{\tau_j}{\sigma^2} \middle| \pmb{\lambda}^{(k-1)},\pmb{y} \right] \\
        & + \text{terms not involving $\lambda_1$ and $\lambda_2$} \\
        &= R(\pmb{\lambda}|\pmb{\lambda}^{(k-1)}) + \text{terms not involving $\lambda_1$ and $\lambda_2$}.
    \end{split}
    \end{equation*}
    Then we find $\lambda_1$ and $\lambda_2$ that maximize $ R(\pmb{\lambda}|\pmb{\lambda}^{(k-1)})$. The maximization procedure can be obtained by using the partial derivatives of $R(\pmb{\lambda}|\pmb{\lambda}^{(k-1)})$
    \begin{equation*}
        \begin{split}
        \frac{\partial R}{\partial \lambda_1}&= \frac{p}{\lambda_1}- \frac{\lambda_1}{4\lambda_2} \sum\limits_{j=1}^p  E\left[\frac{\tau_j}{ \sigma^2} \middle| \pmb{\lambda}^{(k-1)},\pmb{y} \right] \\
        \frac{\partial R}{\partial \lambda_2}&= -\frac{1}{2}\sum\limits_{j=1}^{p} E\left[ \frac{\tau_j}{\tau_j-1}\frac{\theta_j^2}{\sigma^2} \middle| \pmb{\lambda}^{(k-1)},\pmb{y} \right] + \frac{\lambda_1^2}{8\lambda_2^2} \sum\limits_{j=1}^p  E\left[\frac{\tau_j}{\sigma^2} \middle| \pmb{\lambda}^{(k-1)},\pmb{y} \right].
        \end{split}
    \end{equation*}
    Here, the expectations in $Q$ and $R$ are evaluated by using the means of sampled $\pmb{\theta}$, $\pmb{\tau}$, and $\sigma^2$.
    
    Second, the FB approach treats $\pmb{\lambda}$ as unknown model parameters and specify a prior for them. Park and Casella \cite{bayeslasso} suggest to use the gamma prior for the squared value of $l_1$ penalty for BL. Similarly, we assume the $\text{Gamma}(r_1,\delta_1)$ prior on $\lambda_1^2$. Also, we place a $\text{GIG}(\nu_2,\;\psi_2,\;\chi_2)$ prior on $\lambda_2$, which is a conjugate prior \citep{Ben2010}. The full joint posterior density becomes
    \begin{equation*}
	\begin{split}
	\pi\left(\pmb{\theta},\;\pmb{\tau},\,\sigma^2,\lambda_1^2, \lambda_2|X,\;\pmb{y}\right) \propto& L_{NP}\left(\pmb{\theta}\right)\times \prod_{j=1}^{p}\left(\dfrac{\lambda_2}{\sigma^2}\dfrac{\tau_j}{\tau_j-1} \right)^{\frac{1}{2}} \exp\left( -\dfrac{1}{2}\dfrac{\lambda_2}{\sigma^2}\dfrac{\tau_j}{\tau_j-1}\theta_j^2\right)\\
	&\times \prod_{j=1}^{p} \left(\frac{\lambda_1^2}{8\lambda_2\sigma^2}\right)^{\frac{1}{2}}\tau_j^{-\frac{1}{2}}\exp\left(-\tau_j\dfrac{\lambda_1^2}{8\lambda_2\sigma^2}  \right)I\left(\tau_j \in (1,\infty) \right) \\
	& \times \left(\frac{1}{\sigma^2}\right)^{a+1} \exp\left(-\frac{b}{\sigma^2}\right) \times \left(\lambda_1^2\right)^{r_1-1} \exp\left(-\delta_1\lambda_1^2\right) \\
	& \times  \left(\lambda_2\right)^{\nu_2-1} \exp\left\{-\frac{1}{2}\left(\chi_2\frac{1}{\lambda_2}+ \psi_2\lambda_2\right) \right\}.
	\end{split}
	\label{M13}
	\end{equation*}

    The full conditional distributions for the penalty parameters are
    \begin{equation*}
       \begin{split}
        \lambda_1^2 | \pmb{\tau}, \sigma^2, \lambda_2 &\sim \text{Gamma}\left( \frac{p}{2}+ r_1,\sum_{j=1}^p \frac{\tau_j}{8\lambda_2\sigma^2}+\delta_1\right) \\
        \lambda_2 | \pmb{\tau}, \sigma^2, \lambda_1^2, \pmb{\theta} &\sim \text{GIG} \left(\nu = \nu_2, \psi = \sum\limits_{j=1}^p \frac{\tau_j}{\tau_j-1}\frac{\theta_j^2}{\sigma^2} + \psi_2, \chi = \sum\limits_{j=1}^p \frac{\tau_j \lambda_1^2}{4\sigma^2} + \chi_2 \right).
        \end{split}
    \end{equation*}
    
    An alternative prior for $\lambda_2$ to consider is $\text{Gamma}(r_2,\delta_2)$. Then, the full conditional distribution becomes
        \begin{equation*}
       \begin{split}
        \lambda_2 | \pmb{\tau}, \sigma^2, \lambda_1^2, \pmb{\theta} &\sim \text{GIG} \left(\nu = r_2, \psi = \sum\limits_{j=1}^p \frac{\tau_j}{\tau_j-1}\frac{\theta_j^2}{\sigma^2} + 2\delta_2, \chi = \sum\limits_{j=1}^p \frac{\tau_j \lambda_1^2}{4\sigma^2} \right).
        \end{split}
    \end{equation*}
    The gamma distribution is a special case of the GIG distribution family with $\chi=0$. Thus, placing a gamma prior on $\lambda_2$ results on a posterior distribution that follows a GIG distribution.

    \subsection{Variable selection methods}
    	\label{sec:CI}
    	The Bayesian regularization approaches require additional variable selection methods. In most non-Bayesian penalized methods, the estimated coefficients shrink to zero. On the other hand, the outputs of Bayesian models are the posterior distributions, which are not necessarily equal to zero. We present two variable selection methods: Bayesian credible interval and scaled neighborhood criteria.  
	
	First, the Bayesian credible interval can be used to select variables whose credible regions do not include zero for a given probability level $\alpha$ \citep{li2011bayesian}. 
	Lazar \cite{Lazar2003} proves that under standard regularity conditions and as $n \rightarrow \infty$, the posterior EL distribution $\pmb{\theta}(F)$ converges to the normal distribution. 
	Bedoui and Lazar \cite{Adel2020} show that the asymptotic posterior distribution of the BEL version of ridge and lasso regressions follows the normal distribution.
	Lemma~\ref{lemma} shows that the posterior EL distribution of $\pmb{\theta}$ for the elastic net model converges to a normal distribution as $n \rightarrow \infty$. However, we want to note that the Bayesian credible interval may include zero in the presence of strong correlation among explanatory variables because of bimodality of the posterior of the regression parameters \citep{pasanen2015bayesian}.
	
	\begin{lemma} Assume standard regularity conditions, $\nabla \log\left( \pi(\pmb{\theta}) \right) = 0$ and $\nabla \log\left( \pi(X,\pmb{y}| \pmb{\theta})\right) = 0$. The posterior EL distribution of $\pmb{\theta}$ for the Bayesian elastic net converges to the normal distribution with mean $\pmb{m}$ and covariance $\mathit{J_n}$ as $ n \rightarrow \infty$, where
    \[
	\begin{split}
	\mathit{J_n}&=\dfrac{\lambda_2}{\sigma^2}D_{\tau} + \mathit{J}(\pmb{\hat{\theta}}_n),\\
	\pmb{m}&= \mathit{J_n}^{-1} 
	\left( \dfrac{\lambda_2}{\sigma^2}D_{\tau} \pmb{\theta_0} +\mathit{J}(\pmb{\hat{\theta}}_n)\pmb{\hat{\theta}}_n\right),\\
	\end{split}
	\]
	$\pmb{\theta_0}$ is the prior mode, 
	%(Is there a reason why you did not write this term in your CSDA paper and ours? It is in Nicole's paper, but not in your CSDA and ours. If there is a reason, then please remove them!) Are you sure that you did read my paper? :-)  in page 9, section 6 I explicitly refer the reader to check my supplementary material to look for the normality assumption. In my supplementary paper, part 3 page 5, I layout there in details the proof. If you are going to ask me why I did in that way? Nicole and I  agreed that will be the best option. Also, recall that you have modified this part many times and I believe that originally I had written something different than what we have now :-) SO I AM NOT CHECKING THE REST OF THE PROOF UNTIL YOU CHECK THIS,OK? 	
	$\pmb{\hat{\theta}}_n$ is the profile maximum likelihood estimate of $\pmb{\theta}$ and 
	\[
	\mathit{J}(\pmb{\hat{\theta}}_n) = \left[ \dfrac{\partial^2}{\partial \theta_i\partial \theta_j}\sum_{i=1}^{n}\log \left\{ 1+\pmb{\gamma}^T\pmb{x_i}(y_i-\pmb{x_i}^T\pmb{\theta})\right\} \right]_{\pmb{\theta}=\pmb{\hat{\theta}}_n}.
	\]
	Also, $-2\log\left(\pi(\pmb{\theta}|\bm \tau, \sigma^2,\mathit{X},\pmb{y}) \right) \xrightarrow{d} \chi_p^2$ as $ n \rightarrow \infty$.
	\label{lemma}
	\end{lemma}
	
	\begin{proof}
	Under the standard regularity conditions, the prior term is dominated by the likelihood function, and the log posterior distribution of $\pmb{\theta}$ can be expressed using up to quadratic order terms \citep{Bernardo1994},
%The standard regularity conditions enable us to express the log prior distribution of $\pmb{\theta}$ using up to quadratic order terms so that
\begin{equation*}
   -2\log\left(\pi(\pmb{\theta}|\bm \tau, \sigma^2,\mathit{X},\pmb{y}) \right)  \propto (\pmb{\theta}-\pmb{m})^TJ_n^{-1}(\pmb{\theta}-\pmb{m}).
\end{equation*}
%Here, the prior term will be dominated by the likelihood function \citep{Bernardo1994}.
Lazar \cite{Lazar2003} shows that the posterior distribution of $\pmb{\theta}$ converges to normal distribution with mean $\pmb{m}$ and variance $J_n$ as $n\rightarrow \infty$. Then, showing $J_n$ is symmetric and positive semi-definite completes the proof. 

First, the first term of $J_n$, the diagonal matrix $\dfrac{\lambda_2}{\sigma^2}D_{\tau}$, is the minus second derivative of the log prior distribution of $\pmb{\theta}$ in (\ref{M6}) evaluated at $\pmb{\theta}_0$.
Also, the second term of $J_n$, $J(\pmb{\theta})$, can be approximated as Fisher information matrix \citep{Bernardo1994}, which is a positive semi-definite matrix by definition. 
%Also, it is easy to show that 
%$\frac{\partial}{\partial \pmb{\theta}}J(\pmb{\theta})\leq 0$ and 
%$\frac{\partial^2}{\partial \pmb{\theta}\partial \pmb{\theta}^T} \sum_{i=1}^{n}\log \left\{ 1+\pmb{\gamma}^T\pmb{x_i}(y_i-\pmb{x_i}^T\pmb{\theta})\right\}>0$ \textcolor{red}{this is actually <0... I don't know what mistake that I made here... So $J(\pmb{\theta})$ is Hessian matrix of minus empirical likelihood. To this to be positive definite, the minus empirical likelihood needs to be convex. However, log(-theta) is concave...}. 
Therefore, $J_n$ is positive semi-definite because both $\dfrac{\lambda_2}{\sigma^2}D_{\tau}$ and $\mathit{J}(\pmb{\hat{\theta}}_n)$ positive semi-definite. Thus, $-2\log\left(\pi(\pmb{\theta}|\bm \tau, \sigma^2,\mathit{X},\pmb{y}) \right) \xrightarrow{d} \chi_p^2$ as $ n \rightarrow \infty$.
%The rest of the proof follows similar to Theorem of \cite{Lazar2003} and Lemma 1 of \cite{Adel2020}. \textcolor{blue}{do we need to write more?}
	\end{proof}

	%Based on this result, 
	%Thus, by the normality result of the posterior EL distribution of $\pmb{\theta}$ aforementioned and under standard regularity conditions, $-2\log\left(\pi(\pmb{\theta}|\bm \tau, \sigma^2,\mathit{X},\pmb{y}) \right)$ converges in distribution to $\chi_p^2$ as $ n \rightarrow \infty$. 

	%\subsubsection{Scaled Neighborhood}
	Second, the scaled neighborhood criterion uses the posterior probability to select variables \citep{Ben2010}. The variable $\theta_j$ is excluded when the posterior probability $P\left(|\theta_j| \leq \sqrt{\text{Var} (\theta_j \mid \pmb{y})} \mid \pmb{y}\right)$ is greater than $\eta$ for a given threshold $\eta$. %In our paper, we select variables using the scaled neighborhood criterion with threshold level $\eta=0.5$. 

    %````````````````````````````````````````````````%
	% Simulation studies                             %
	%````````````````````````````````````````````````%
    \section{Simulation studies}
    \label{sec:sim}
    We conduct Monte Carlo simulations to compare the performance of the proposed Bayesian elastic net based on the empirical likelihood (BEN-EL) model with four different models: BEN, BL, EN, and lasso applied to the least absolute deviation regression (LADL). LADL is known to be robust to heavy tail errors and outliers \citep{wang2007robust}.
    We generate data from the linear model
    \begin{equation*}
        \pmb{y}=\pmb{X}\pmb{\theta}+\epsilon.
    \end{equation*}
    Note that BEN-EL and LADL assume that the mean and the median of errors are zero, respectively. On the other hand, BEN, BL, and EN assume normal errors.  
    
    \subsection{Prediction performance}
    \label{subsec:sim.pred}
    \subsubsection{Simulation 1}
    
    In Simulation 1, we set $\pmb{\theta}=(3,1.5,0,0,2,0,0,0)^T$. The explanatory variables $X_{ij}$'s are generated from the standard normal distribution with pairwise correlations $\rho(x_i,x_j)=0.5^{|i-j|}$ for all $i,j\in\{1,...,8\}$. We generate training data sets with three different sizes ($n=50, 100, 200$). For each setting, we generate test data sets of size $400$.
    The error is generated from three different distributions: 1) normal distribution, $\epsilon \sim N(0,3^2)$, 2) non-normal distribution, $\epsilon \sim N(-3,1)$ or $\epsilon \sim N(3,1)$ with probability of 0.5, respectively, and 3) skew Student $t$ distribution proposed by Fern{\'a}ndez and Steel \cite{fernandez1998bayesian}, $\epsilon \sim ST(\nu=30,\xi=1.5)$ with mean 0 and standard deviation 3. We use the R package fGarch \citep{fGarch} to sample from the skew Student $t$ distribution. The sampling procedures are repeated 100 times for each simulation setting. 
    %Under each combination of simulation settings, 100 data sets are generated.

    The shrinkage parameters $\lambda_1$ and $\lambda_2$ are estimated using the EB method for BEN-EL, BEN, and BL methods. For BEN-EL, the hyperparameters $a=10$ and $b=10$ are used. For the bisection step size tuning, the initial step size $\omega=0.5$ and tolerances $\iota_l=\iota_u=0.05$ are used.
    For BEN and BL, we use MCMC with 2,000 burn-ins and 12,000 iterations.
    %where their initial values are set equal to 0.1. 
    The optimal value for $\lambda_2$ in EN is selected from $\{10^n  \mid  n = -2,-1,0,1,2,3 \}$. 
    
    In all cases, the simulated HMC chains meet the convergence criteria $\widehat{R}<1.01$. The trace plots and histograms of estimated $\pmb{\theta}$ by BEN-EL for one dataset of skew Student $t$ errors and $n=50$ are given in Figures~\ref{sfig:sim1theta} and \ref{sfig:sim1hist} in Appendix~\ref{appendix:a}.
    
        \begin{table}[!ht]
        \centering
        \tbl{The median of mean squared prediction errors (MMSPE) along with the standard errors (SE) of BEN-EL, BEN, BL and EN methods for Simulation 1. The smallest MMSPEs are marked in bold.}
        {
\begin{tabular}{clrrrrr}
\hline
\multicolumn{1}{c}{Error} & \multicolumn{1}{l}{} & \multicolumn{5}{c}{MMSPE (SE)}                                                                        \\ \cline{3-7} 
                     &                       & \multicolumn{1}{c}{BEN-EL} & \multicolumn{1}{c}{BEN} & \multicolumn{1}{c}{BL} & \multicolumn{1}{c}{EN}
                     & \multicolumn{1}{c}{LADL}
                     \\ \hline
$N(0,3^2)$             & $n=50$                & 10.24 (0.18)              & 11.33 (0.25)            & \textbf{10.00 (0.16)}           & 10.43 (0.15) & 11.07 (0.18)          \\
                     & $n=100$               & 9.61 (0.10)               & 11.65 (0.19)            & \textbf{9.55 (0.10)}            & 9.73 (0.10)         &  9.96 (0.15) \\
                     & $n=200$               & 9.13 (0.07)               & 11.65 (0.10)            & \textbf{9.10 (0.08)}            & 9.17 (0.07)         & 9.36 (0.12)   \\ \hline
$N(3,1)$ & $n=50$                & \textbf{10.80 (0.12)}              & 12.33 (0.25)            & 10.91 (0.12)           & 11.20 (0.17)  & 14.37 (0.29)         \\
       or              & $n=100$               &  \textbf{10.49 (0.11)}                         & 12.36 (0.21)            & 10.49 (0.11)           & 10.68 (0.09) & 13.09 (0.18)          \\
$N(-3,1)$ & $n=200$               &  10.29 (0.05)                         & 12.66 (0.13)            & \textbf{10.27 (0.06)}           & 10.34 (0.07)   & 12.52 (0.10)        \\ \hline
Skew & $n=50$                & \textbf{90.20 (0.99)}              & 91.38 (1.30)            & 91.83 (1.22)           & 93.93 (1.30)   & 92.29 (0.76)          \\
        Student $t$            & $n=100$               &  86.31 (1.26)                        & 86.77 (1.07)            & \textbf{86.09 (1.12)}           & 88.50 (1.48) & 88.15 (1.08)        \\
& $n=200$               &  \textbf{83.56 (0.80)}                         & 83.95 (0.68)            & 83.66 (0.76)           & 84.51 (0.89)  & 84.81 (1.05)         \\ \hline
\end{tabular}
}
\label{tb:sim1.rmse}
\end{table}
    
    BEN-EL outperforms the other methods when the number of observations is small ($n=50,100$) and the normality assumption in errors is violated.
    Table~\ref{tb:sim1.rmse} presents the median of mean squared prediction error (MMSPE) and the standard error (SE) results based on Simulation 1. For Bayesian methods, the explanatory variables are selected based on the scaled neighborhood criterion with $\eta>0.5$ in computing MMSPE. The SE is computed as a standard deviation of 1,000 bootstrap samples of mean squared prediction error values. %When the error follows the normal distribution, the MMSPE and SE values for the BEN-EL, BL, and EN methods are quite similar, whereas they are higher for the BEN approach.  
    %The BEN-EL and BL methods outperform the BEN and EN methods and are quite similar when $n=200$.
   %and based on the MSPE and the SE values, the BEN-EL method outperforms the BEN approach whereas it performs quite similar to BL and EN methods. 
   
   \begin{table}[!ht]
   \centering
   \tbl{The empirical frequency (\%) that the regression coefficient is dropped for Simulation 1 with errors from $N(3,1)$ or $N(-3,1)$. For BEN-EL, BEN, and BL, the scaled neighborhood criterion with $\eta>0.5$ is used.  }
   {
\begin{tabular}{lcrrrrrrrr}
\hline
                         & \multicolumn{1}{l}{} & \multicolumn{8}{c}{Empirical frequency of exclusion}                                                                                                                                                                                                                       \\ \cline{3-10} 
                         & \multicolumn{1}{l}{} & \multicolumn{1}{c}{$\theta_1$} & \multicolumn{1}{c}{$\theta_2$} & \multicolumn{1}{c}{$\theta_3$} & \multicolumn{1}{c}{$\theta_4$} & \multicolumn{1}{c}{$\theta_5$} & \multicolumn{1}{c}{$\theta_6$} & \multicolumn{1}{c}{$\theta_7$} & \multicolumn{1}{c}{$\theta_8$} \\ \hline
\multirow{3}{*}{$n=50$}  & BEN-EL                & 0                             & 0                             & 58                            & 69                            & 0                             & 65                            & 75                            & 64                            \\
                         & BEN                  & 0                             & 0                             & 34                            & 39                            & 0                             & 53                            & 80                            & 76                            \\
                         & BL                   & 0                             & 3                             & 69                            & 76                            & 0                             & 77                            & 82                            & 75  
                          \\
                         & LADL                   & 0                             & 18                             & 70                            & 76                            & 9                             & 86                            & 90                            & 82\\ \hline
\multirow{3}{*}{$n=100$} & BEN-EL                & 0                             & 0                             & 70                            & 61                            & 0                             & 59                            & 72                            & 69                            \\
                         & BEN                  & 0                             & 0                             & 13                            & 15                            & 0                             & 34                            & 73                            & 74                            \\
                         & BL                   & 0                             & 0                             & 80                            & 73                            & 0                             & 75                            & 82                            & 77  \\
                         & LADL                   & 0                             & 12                             & 76                            & 73                            & 5                             & 81                            & 88                            & 84                           \\ \hline
\multirow{3}{*}{$n=200$} & BEN-EL                & 0                             & 0                             & 56                            & 66                            & 0                             & 68                            & 74                            & 62                            \\
                         & BEN                  & 0                             & 0                             & 2                             & 3                             & 0                             & 17                            & 68                            & 82                            \\
                         & BL                   & 0                             & 0                             & 69                            & 69                            & 0                             & 75                            & 82                            & 75   \\
                         & LADL                   & 0                             & 6                             & 77                            & 71                            & 0                             & 81                            & 88                            & 84                          \\ \hline
\end{tabular}}
\label{tb:sim1.num}
\end{table}

Table~\ref{tb:sim1.num} reports the number of exclusion of each variable for 100 simulated data sets for BEN-EL, BEN, BL, and LADL. The variable selection results for EN are not included because all the estimated coefficients are nonzero. LADL performs the best in dropping variables with zero coefficients ($\theta_3$, $\theta_4$, $\theta_6$, $\theta_7$, and $\theta_8$) but some nonzero coefficients ($\theta_1$, $\theta_2$, and $\theta_5$) are also excluded. Among the Bayesian methods, BL best identifies the zero coefficients. However, the MMSPEs for BL are larger than BEN-EL for $n=50,100$. This implies that BEN-EL makes more accurate estimates even though their zero variables exclusion rate is worse than BL. %When it comes to variable selection, BEN performs poorly as the number of observation increases for simulation 1 data.

%and definitely, in general, does better than the BEN method. NOT SURE Maybe BEN-EL is better in estimating coefficients, because it considers the correlation structure unlike BL? - languages are not tailored yet...I will come back later or you are welcome to interpret. }

    \subsubsection{Simulation 2}
    Simulation 2 dataset has a more complex correlation structure than Simulation 1. 
    First, we generate $Z_1$, $Z_2$, and $Z_3$ independently from $N(0,1)$. Then, we add normal errors so that $x_i=Z_1+\upsilon_i$  for $i=1,\ldots,5$, $x_i=Z_2+\upsilon_i$ for $i=6,\ldots,10$, $x_i=Z_3+\upsilon_i$ for $i=11,\ldots,15$, and $x_i \sim N(0,1)$ for $i=16,\ldots,30$, where $\upsilon_i \sim N(0,0.01)$ for $i=1,\ldots,15$. As a result, the variables in the first three groups are highly correlated within each group. We set the true parameters of the linear model as
    \begin{equation*}
        \pmb{\theta}=(\underbrace{3,\cdots,3}_{5},\underbrace{3,\cdots,3}_{5},\underbrace{3,\cdots,3}_{5},\underbrace{0,\cdots,0}_{15})^T.
    \end{equation*}

    We generate error $\epsilon$ using three different distributions: 1) normal distribution, $\epsilon \sim N(0,15^2)$, 2) Student $t$ distribution with degrees of freedom 3, $\epsilon \sim 10 \times t_3 $, and 3) skew Student $t$ distribution, $\epsilon \sim ST(\nu=30,\xi=1.5)$. 
    The normal and $t$ distributions are symmetric and have similar variances, but the $t$ distribution has a thicker tail than the normal distribution. On the other hand, the skew Student $t$ distribution is right-skewed.  
    %That is, the normality assumption is violated in the second settings. 
    We use training data sets with two different sizes ($n=100,200$) and fix the size of testing data sets to $400$. We generate 100 data sets for each simulation setting.
    %Also, we increase the size of training and testing sets to 200 and 200 under the same setting of Simulation 1.
    For the other simulation parameters, we use the same setting used in Simulation 1.
    
    For BEN-EL, the simulated HMC chains converge with $\widehat{R}<1.01$ in all simulation settings. We also present the trace plots and histograms of four nonzero coefficients $(\theta_1,\theta_2,\theta_3,\theta_4)$ and four zero elements $(\theta_{16},\theta_{17},\theta_{18},\theta_{19})$ estimated by BEN-EL for one dataset of the skew Student $t$ errors and $n=100$ in Figures~\ref{sfig:sim2theta} and \ref{sfig:sim2hist} in Appendix~\ref{appendix:a}.
    %\textcolor{red}{add trace plot and histogram}
    
\begin{table}[!ht]
\centering
\tbl{The median of mean squared prediction errors (MMSPE) along with the standard errors (SE) of BEN-EL, BEN, BL, EN, and LADL methods for Simulation 2. The smallest MMSPEs are marked in bold.}
{
\begin{tabular}{clrrrrr}
\hline
\multicolumn{2}{l}{\multirow{2}{*}{}}                      & \multicolumn{5}{c}{MMSPE (SE)}                                                                                                       \\ \cline{3-7} 
\multicolumn{2}{l}{}         & \multicolumn{1}{c}{BEN-EL}                & \multicolumn{1}{c}{BEN}                 & \multicolumn{1}{c}{BL} & \multicolumn{1}{c}{EN} & \multicolumn{1}{c}{LADL} \\ \hline
\multirow{2}{*}{Normal}      & \multicolumn{1}{c}{$n=100$} & \textbf{281.98 (3.70)}  & 291.44 (4.50)                           & 295.28 (4.16)          & 309.52 (5.36)  & 332.72 (3.42)        \\
                             & \multicolumn{1}{c}{$n=200$} & 255.06 (2.58)                            & \textbf{254.12 (3.02)} & 258.17 (3.35)          & 263.62 (3.42)   & 323.32 (3.73)        \\ \hline
\multirow{2}{*}{Student $t$} & $n=100$                     & \textbf{334.10 (10.75)} & 344.72 (7.61)                           & 345.03 (7.79)          & 380.51 (12.33)    & 374.70 (5.78)     \\
                             & $n=200$                     & 313.37 (6.77)                            & \textbf{302.26 (6.46)} & 307.12 (6.62)          & 311.59 (6.13) & 351.81 (7.51)         \\ \hline
Skew  & $n=100$                     & \textbf{289.95 (4.03)} & 297.96 (4.43)                           & 301.23 (4.64)          & 318.13 (8.88)   & 344.05 (2.83)      \\Student $t$
                             & $n=200$                     & \textbf{253.97 (3.06)}                            & 255.24 (2.74) & 256.02 (2.97)          & 262.62 (3.52)      & 310.68 (4.33)    \\ \hline 
\end{tabular}}
\label{tb:sim2.rmse}
\end{table}

The simulation results suggest that BEN-EL outperforms when the sample size is small or the error follows the asymmetric distribution.
Table~\ref{tb:sim2.rmse} reports the MMSPE and SE results of Simulation 2. 
In all cases, the Bayesian methods (BEN-EL, BEN, and BL) perform better than the non-Bayesian methods (EN and LADL), and the Bayesian EN methods (BEN-EL and BEN) perform better than the BL. This corresponds to the findings that the EN-based methods perform better than the lasso-based methods under the complex correlation structure \citep{Zou2005}.
For the symmetric error distributions (normal and Student $t$), BEN-EL provides the smallest MMSPE and SE values compared to BEN when the sample size is small. However, BEN-EL performs the best under the skewed error distribution regardless of the sample size.
%Bayesian EN-based methods (BEN-EL and BEN) perform well compared to the lasso-based method (BL) and frequentist method (EN). 
%Bayesian methods performs better than the frequentist method EL. Among Bayesian methods, EN-based methods (BEN-EL and BEN) achieve smaller MMSPE than lasso-based method (BL). For both normal and $t$ distribution, BEN-EL outperforms when the training set size is small ($n=100$) whereas BEN ourperforms when the set size is large ($n=200$). 
%When the error follows the normal distribution, the MSPE and SE values in the BEN-EL, BEN, and BL methods are quite similar where it is worse in the EN approach. On the other hand, the BEN-EL achieves the smallest MMSPE and SE when the error follows the Student $t$ distribution.
%In the casewhen the error follows $t$ distribution, and based on the MSPE and the SE values, the BEN-EL method outperforms the BEN approach whereas it performs quite similar to BL and EN methods. 

Table~\ref{tb:sim2.var} presents the variable selection results. For the Bayesian methods, variables are selected using the scaled neighborhood criterion with $\eta>0.5$. The variable selection results for EN are not reported because its estimated coefficients are nonzero. 
%The BEN-EL approach is more conservative in excluding variables compared to the BEN and BL when it comes to the variables selection. 
BEN-EL tends to keep more variables than BEN and BL for both nonzero and zero coefficients. % and provides better performance when the sample size is small. 
%That is, the BEN-EL method performs well when data provide little information. 
BEN and BL perform poorly in keeping nonzero variables when the sample size is small, but they improve as the sample size increases. LADL shows the highest exclusion rate for zero coefficients but removes many nonzero coefficients.

\begin{table}[!ht]
\tbl{The empirical frequency (\%) that the 15 nonzero and 15 zero regression coefficients are dropped for Simulation 2. For Bayesian methods, the scaled neighborhood criterion with $\eta>0.5$ is used. }
{
\begin{tabular}{ccrrrrrrrr}
\hline
\multicolumn{1}{l}{}         & \multicolumn{1}{l}{} & \multicolumn{8}{c}{Empirical frequency of exclusion}                                                                                                                             \\ \cline{3-10} 
                             &                      & \multicolumn{2}{c}{BEN-EL}                              & \multicolumn{2}{c}{BEN}                                & \multicolumn{2}{c}{BL}  
                             & \multicolumn{2}{c}{LADL}     \\ \cline{3-10} 
                             &                      & \multicolumn{1}{c}{Nonzero} & \multicolumn{1}{c}{Zero} & \multicolumn{1}{c}{Nonzero} & \multicolumn{1}{c}{Zero} & \multicolumn{1}{c}{Nonzero} & \multicolumn{1}{c}{Zero} & \multicolumn{1}{c}{Nonzero} & \multicolumn{1}{c}{Zero}\\ \hline
Normal                       & $n=100$              & 22.26                     & 63.33                  & 34.60                     & 74.07                  & 42.33                     & 87.53      &   71.20 & 92.93                \\
                             & $n=200$              & 5.47                      & 68.47                  & 5.60                      & 69.27                  & 9.87                      & 82.20     & 48.87 & 94.40       \\ \hline
\multirow{2}{*}{Student $t$} & $n=100$              & 23.73                     & 59.13                  & 41.73                     & 78.40                  & 48.93                     & 88.47   &64.67 & 94.40                \\
                             & $n=200$              & 10.53                     & 58.80                  & 12.13                     & 70.07                  & 16.60                     & 83.20         & 35.93 & 93.87         \\ \hline
Skew & $n=100$              & 21.53                     & 61.07                  & 32.53                     & 76.00                  & 38.67                     & 87.53          & 72.07 & 94.00        \\
              Student $t$               & $n=200$              & 5.40                     & 65.47                  & 5.67                     & 66.53                  & 8.40                     & 81.07            & 45.67 & 92.73\\ \hline
\end{tabular}
}
\label{tb:sim2.var}
\end{table}

    \subsection{Sensitivity of Hyperparameters}
    
    We examine the sensitivity of the hyperparameters of the penalty parameters $\lambda_1$ and $\lambda_2$ in BEN-EL for the FB approach. 
    Li and Lin \cite{Ben2010} suggest that the BEN posterior samples may be heavily dependent on the hyperparameters on the penalty parameters based on their experience. Still, it has not been investigated for BEN-EL.
    Therefore, we investigate how the priors affect posterior inferences by changing hyperparameters: 1) $(r_1,\delta_1)$ of the gamma prior for $\lambda_1^2$ and 2) $(\psi_2, \chi_2)$ of the GIG prior for $\lambda_2$. In both simulations, we use one data set from Simulation 1 with $n=50$ and normal errors $N(0,3^2)$. For the gamma prior, 1,600 combinations of the shape and rate parameters $r_1$ and $\delta_1 \in \{ 0.25, 0.5, \ldots, 9.75, 10 \}$ are used. For the GIG prior, we fix $\chi_2=1$ and use 1,640 combinations of $\nu_2 \in \{-5, -4.75, \ldots ,4.75, 5\}$ and $\psi_2 \in \{ 0.25, 0.5, \ldots , 9.75, 10 \}$. 
    The shape and rate parameters $r_1$ and $\delta_1$ for the gamma prior and $\nu_2$ and $\chi_2$ for the GIG prior affect in the opposite directions for nonzero and zero coefficients. For example, as $r_1$ and $\nu_2$ increase and $\delta_1$ and $\chi_2$ decrease, the estimated nonzero coefficients will decrease, whereas the estimated zero coefficients will increase. 
    
    The step size $\omega$ is estimated by the proposed bisection algorithm using the same input parameters used in Simulation 1. We run four HMC chains of 2,000 iterations with 1,000 burn-ins. In all cases, the split-$\widehat{R}$'s are less than 1.01 and the HMC chains converge.

\begin{figure}[!ht]
\centering
\includegraphics[width=1\textwidth]{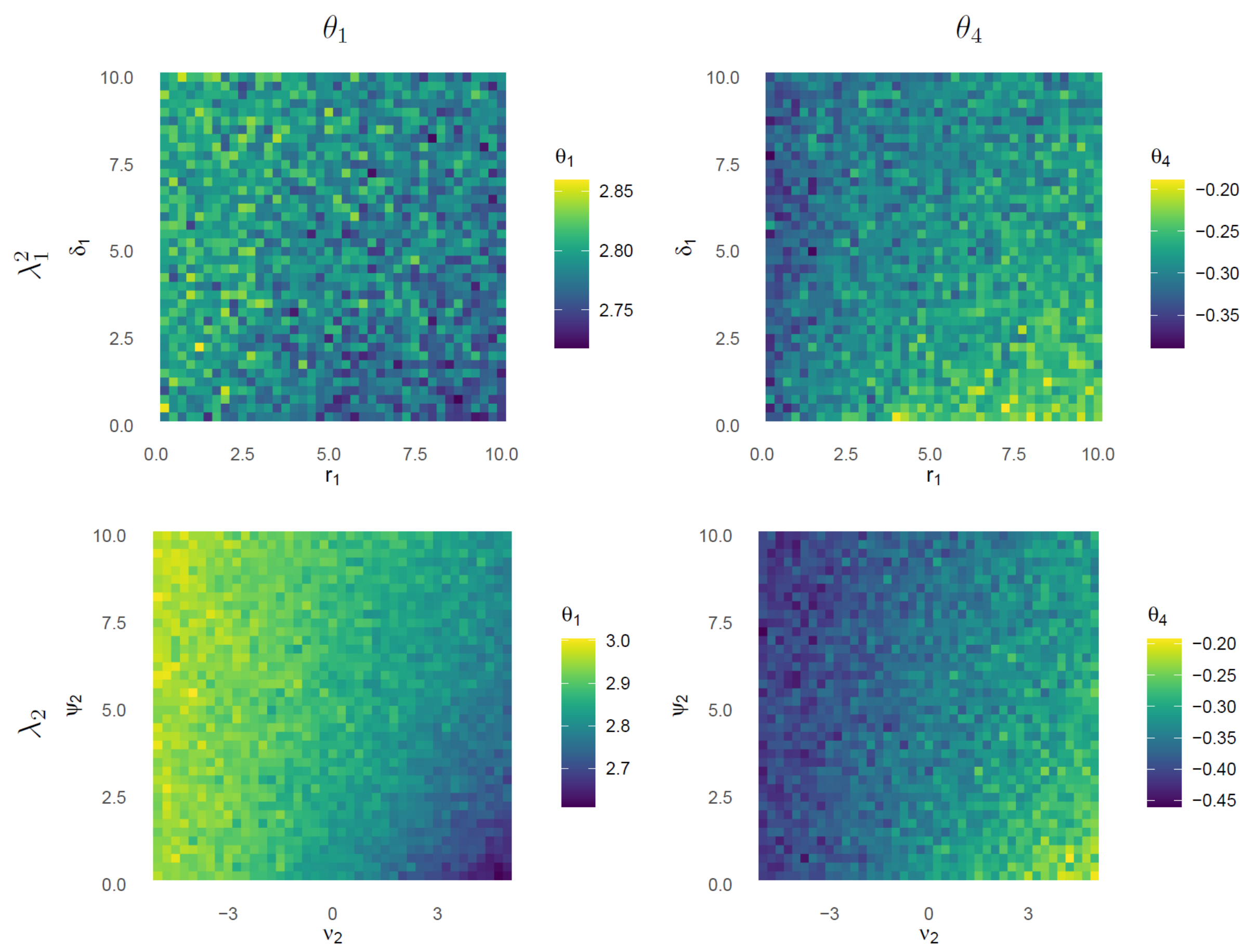}
\caption{Estimated $\theta_1$ (first column) and $\theta_4$ (second column) with respect to the hyperparameters of the gamma prior of $\lambda_1^2$ (first row) and the GIG prior of $\lambda_2$ (second row).}
\label{fig:senstivity}
\end{figure}

The estimated coefficients are affected by choice of the hyperparameters, but the variability is not large enough to interrupt overall inference. Figure~\ref{fig:senstivity} shows heatmaps of estimated coefficients of $\theta_1$ and $\theta_4$ according to different combinations of the hyperparameters.
The estimated coefficients are the median of the posteriors of $\theta_1$ and $\theta_4$ whose true values are 3 and 0, respectively. 
We see the estimated coefficients vary according to the hyperparameters, but they are close to the true coefficients.
For example, the estimated $\theta_1$ differs only up to 0.15 and 0.4 for the various combinations of the hyperparameters of $\lambda_1^2$ and $\lambda_2$, respectively. 
%The GIG hyper-hyperpriors of $\lambda_2$ , although it is difficult to make a direct comparison to the gamma hyper-hyperpriors of $\lambda_1^2$ due to the scale differences.

    \section{Air pollution data analysis}
    \label{sec:real}

    	We use the air pollution to mortality data \citep{mcdonald1973instabilities}. 
	The data set includes the mortality rate ($\pmb{y}$) and 15 explanatory variables of weather, socioeconomics, and air pollution in 60 metropolitan areas of the United States from 1959 to 1961. See Table~\ref{stab:air} in Appendix~\ref{appendix:a} for a description of the variables. 
	
\begin{figure}[!ht]
\centering
\includegraphics[width=0.6\linewidth]{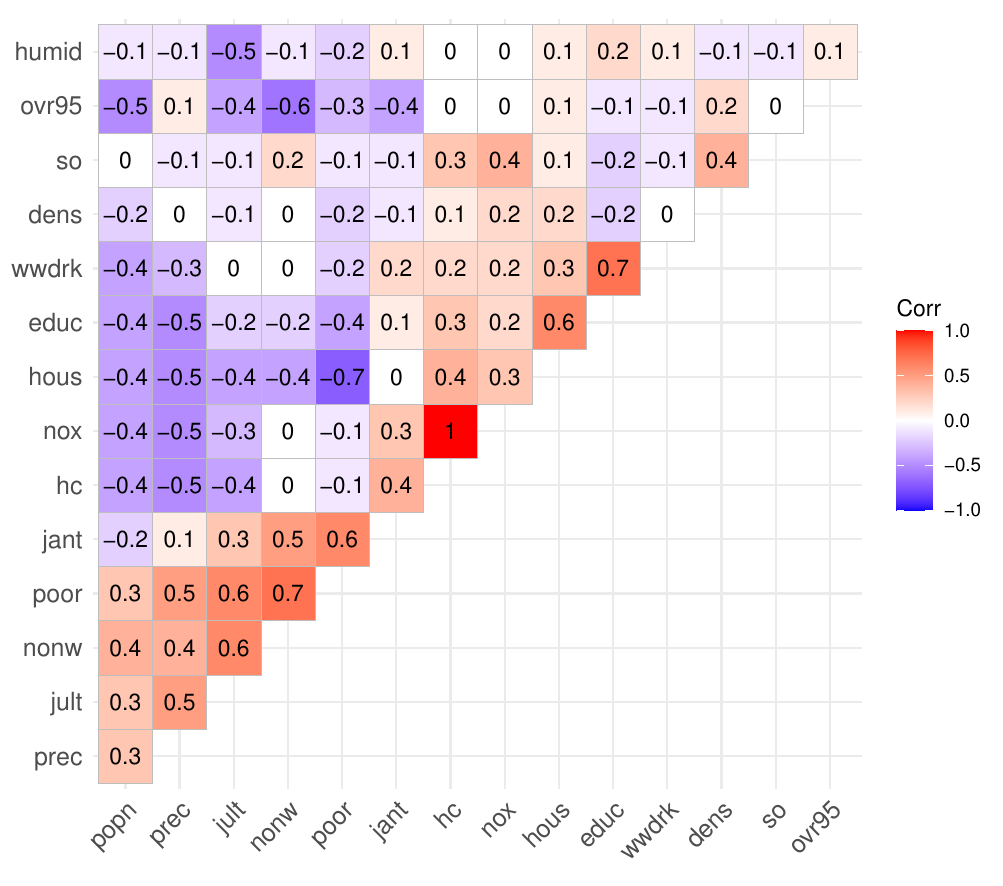} 
\caption{The correlation matrix of the air pollution data.}
\label{sfig:corr}
\end{figure}

	The air pollution set has a complicated correlation structure. The correlation plot is presented in Figure~\ref{sfig:corr}. For example, the nitroc oxide level $nox$ is highly correlated with other air pollution variable hydrocarbon pollution $hc$, but also correlated with the weather variable $prec$ and population variable $popn$. 
	
	We randomly split the data into a training set with 30 observations and a test set of 30 observations. 
	The five models, BEN-EL, BEN, BL, EN, and LADL, are conducted using the training set, and the prediction errors are obtained using the test set. For Bayesian methods, the EB approach is used to select the penalty parameters, and the Bayesian credible interval with the probability level $\alpha=0.5$ is used to select the variables. We repeat the procedure 100 times.
	
	Table~\ref{tb:airpollution} reports the mean prediction error and standard deviation of the five models. BEN-EL yields the smallest mean prediction error, but its standard deviation is the second largest. The relatively large standard deviation of BEN-EL is due to one large prediction error. BEN-EL has the smallest median, mean, the first quartile (Q1), and the third quartile (Q3) than the other methods. Figure~\ref{sfig:box} in Appendix~\ref{appendix:a} shows boxplots of prediction errors of BEN-EL, BEN, BL, LADL. The results imply that the proposed method outperforms the other methods when data have a complex correlation structure and a small number of observations.

	  \begin{table}[!ht]
	  \tbl{The mean prediction errors along with the standard errors (SE) of BEN-EL, BEN, BL and EN methods for air pollution data analysis. The smallest mean prediction error is marked in bold.}
{
\begin{tabular}{rrrrr}
\hline
 \multicolumn{5}{c}{Mean Prediction Error (SE)}                                                                        \\ \hline
                     \multicolumn{1}{c}{BEN-EL} & \multicolumn{1}{c}{BEN} & \multicolumn{1}{c}{BL} & \multicolumn{1}{c}{EN}
                     & \multicolumn{1}{c}{LADL}
                     \\ \hline
 \textbf{1950 (679)}    & 2118 (518)     & 2133 (665)   & 3044 (2903) & 1974 (620) 
                      \\ \hline
\end{tabular}}
\label{tb:airpollution}
\end{table}

%````````````````````````````````````````````````%
% Conclusion                                     %
%````````````````````````````````````````````````%
\section{Conclusion}
\label{sec:conclusion}
We propose a new Bayesian approach for EN based on EL. Accordingly, the profile EL for the linear model replaces the ordinary likelihood function in the Bayesian setting. %\sout{The resulting posterior EL distribution lacks a closed form, and the implementation of even standard MCMC methods is challenging. We use the Hamiltonian Monte Carlo algorithm (also known as hybrid Monte Carlo).}
We use the HMC algorithm because the resulting posterior EL distribution lacks a closed form, and the implementation of even standard MCMC methods is challenging.
A new HMC parameter tuning algorithm based on the bisection method is proposed for BEL.
Simulation studies and the air pollution data application show that our approach outperforms the other methods when the sample size is small, data are correlated, and error distributions are misspecified. 
%\sout{However, the Bayesian elastic net approach has the best prediction accuracy, according to the MSPE and SE values.However, BEN-EL outperforms BEN in the variable selection.}
%Moreover, it appears that BEN-EL is the most robust approach when it comes to variable selection, including when the data distributions are misspecified.

BEL methods share the same limitations of EL in the sense that the number of variables cannot increase as the sample size decreases. The reason is that the estimating equations are included in the convex hull to maximize the profile EL ratio. However, $X^TX$ is not of full rank when $p>n$ and hence it is non-invertible. To address this shortcoming, several approaches have been proposed, such as penalized EL \citep{lahiri2012} and double penalization \citep{chang2018}. Thus, it might be worth a formal investigation whether the proposed method could be extended for $p>n$ case for future work.

Another interesting topic is implementing the NUTS algorithm of \citep{hoffman2014no} to the BEL framework. NUTS automatically tunes leapfrog steps and step sizes and has been used as a default sampler in popular MCMC software such as Stan \citep{stan} and PyMC3 \citep{salvatier2016probabilistic}. However, the proper introduction of NUTS into BEL has not been studied. We observe that HMC chains often diverge when the NUTS algorithm is used for BEN-EL. Also, BEL is not supported by most MCMC libraries because EL cannot be used as a likelihood function. Developing a BEL computational pipeline compatible with Stan or PyMC3 for practitioners will be of further interest.

\section*{Disclosure statement}

No potential conflict of interest was reported by the authors.

\section*{Data availability statement}
	All data used in simulation studies are generated randomly and the air pollution data are obtained from McDonald et al. \cite{mcdonald1973instabilities}. The R code used to generate, import, and analyze the data used in this paper is publicly available at the URL: \url{https://github.com/chulmoon/BEN-EL}.
	
%An unnumbered section, e.g.\ \verb"\section*{Disclosure statement}", may be used to declare any potential conflict of interest and included \emph{in the non-anonymous version} before any Notes or References, after any Acknowledgements and before any Funding information.

\bibliographystyle{tfnlm}
\bibliography{reference_abb}

\appendix

\section{}
\label{appendix:a}

\begin{figure}[!ht]
\centering
\includegraphics[width=0.35\linewidth]{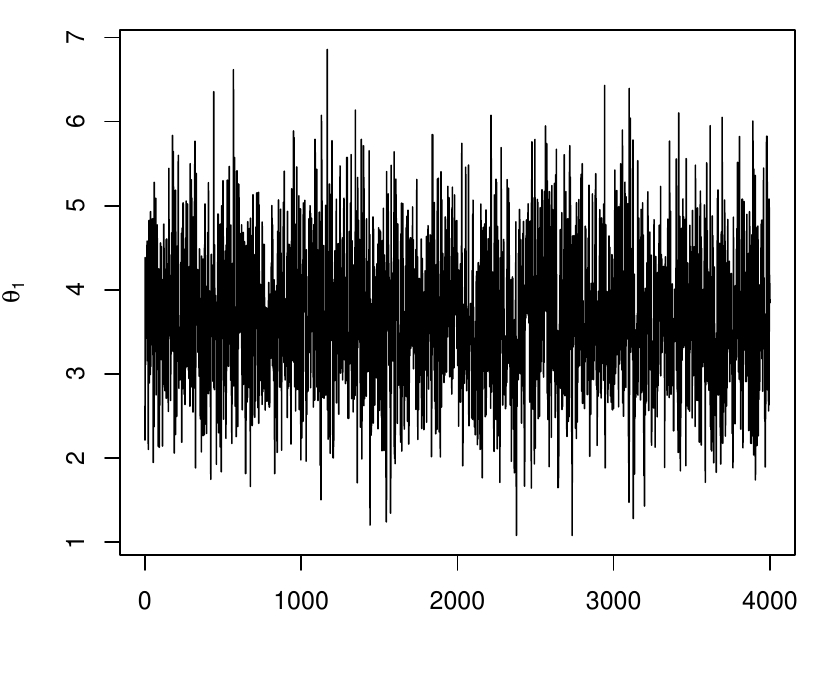} 
\includegraphics[width=0.35\linewidth]{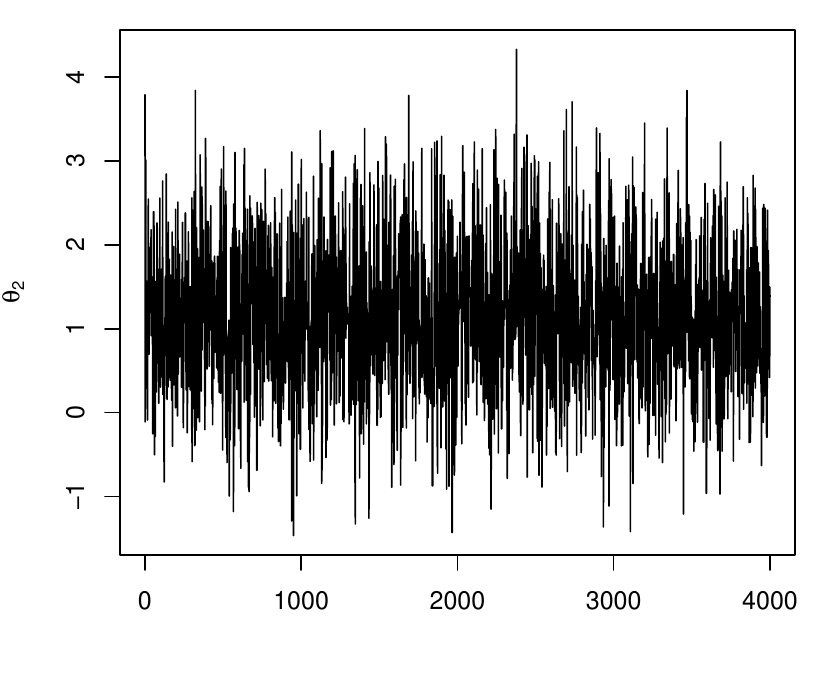} 
\includegraphics[width=0.35\linewidth]{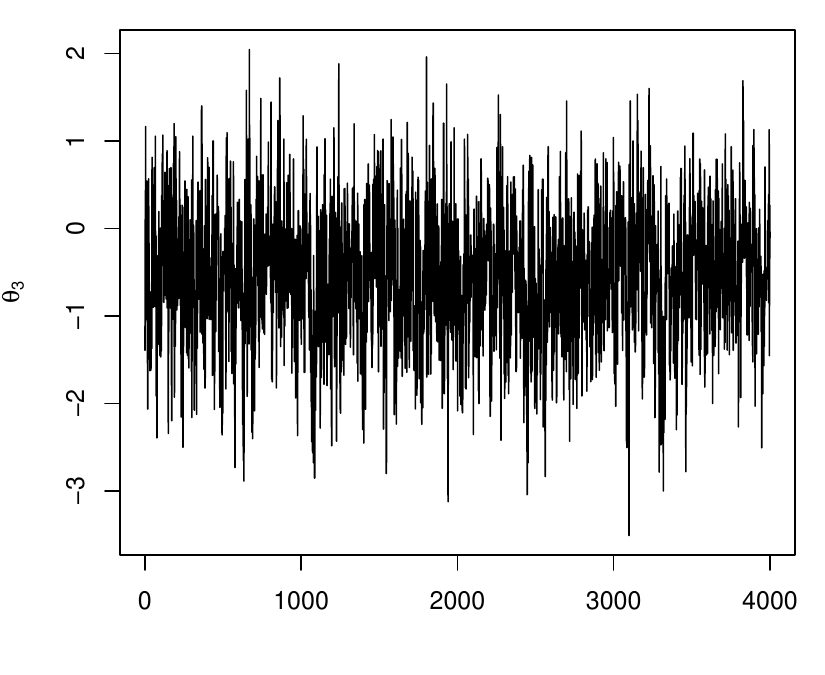} 
\includegraphics[width=0.35\linewidth]{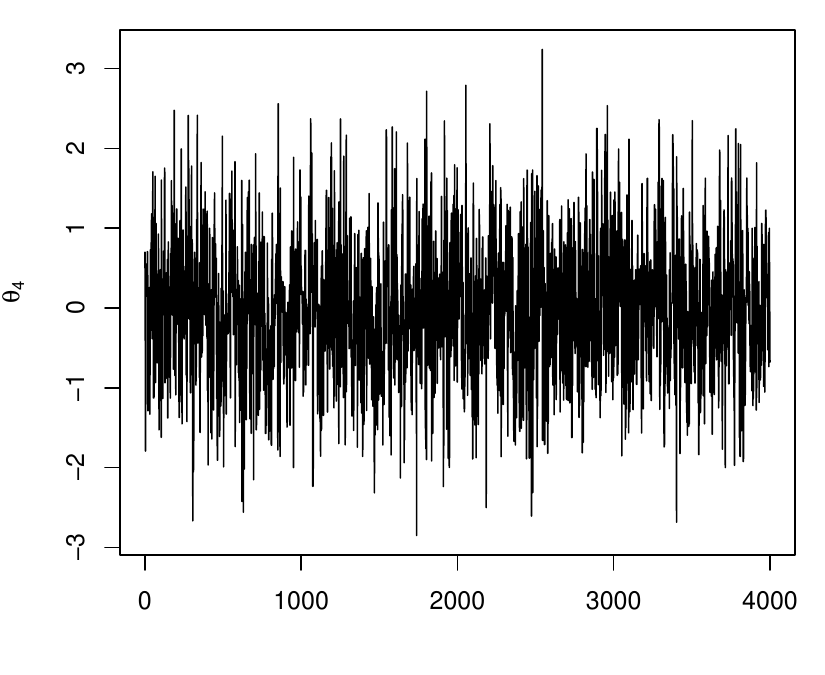} 
\includegraphics[width=0.35\linewidth]{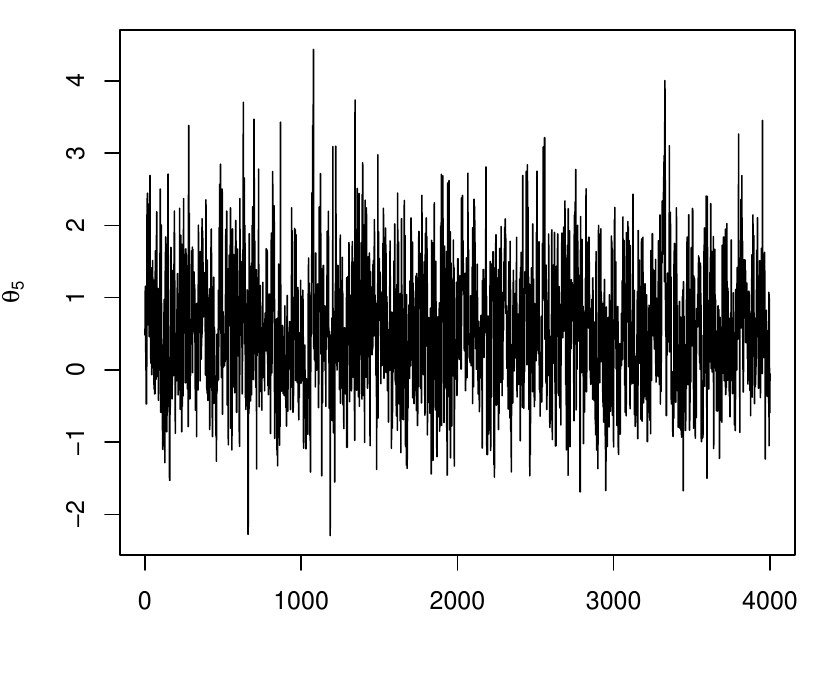} 
\includegraphics[width=0.35\linewidth]{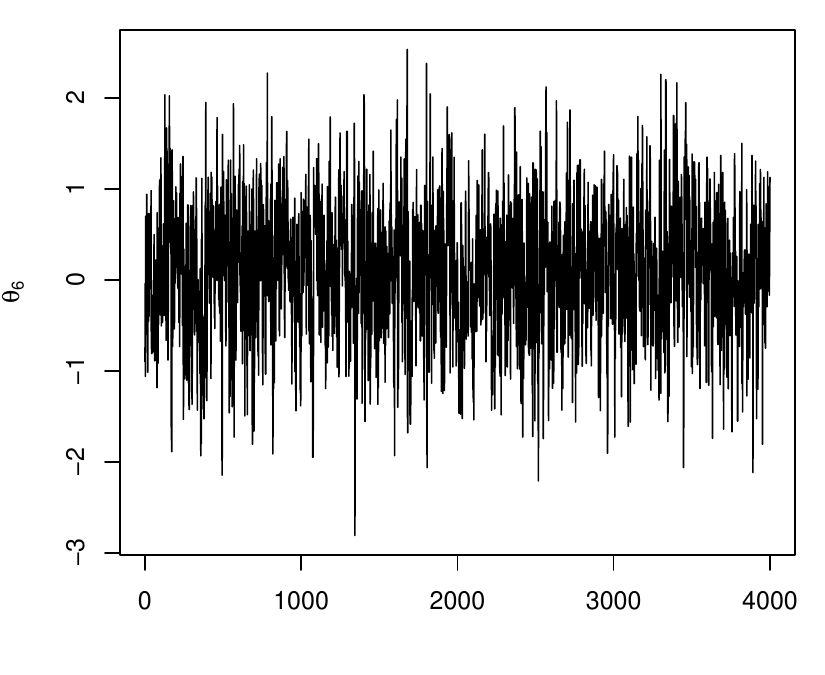} 
\includegraphics[width=0.35\linewidth]{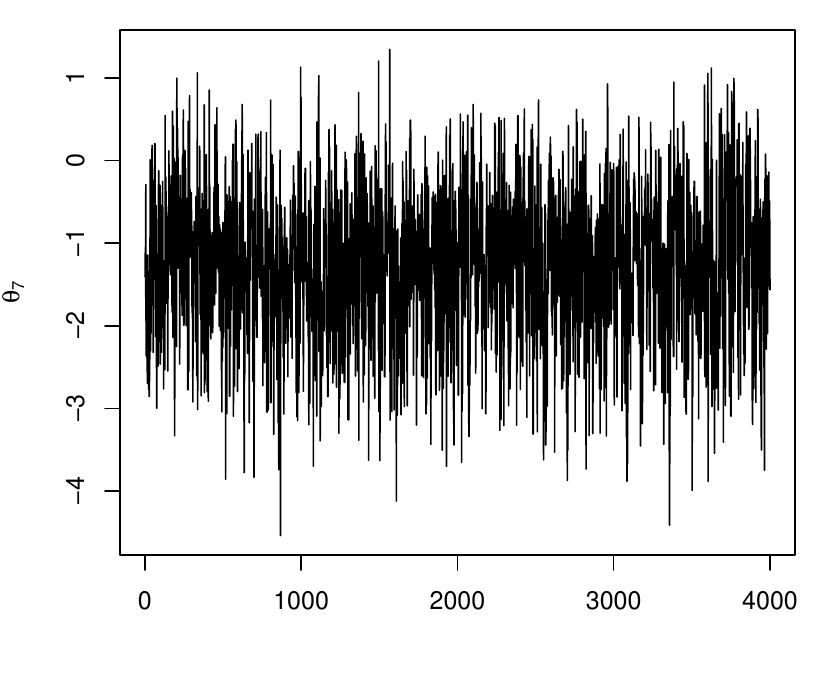} 
\includegraphics[width=0.35\linewidth]{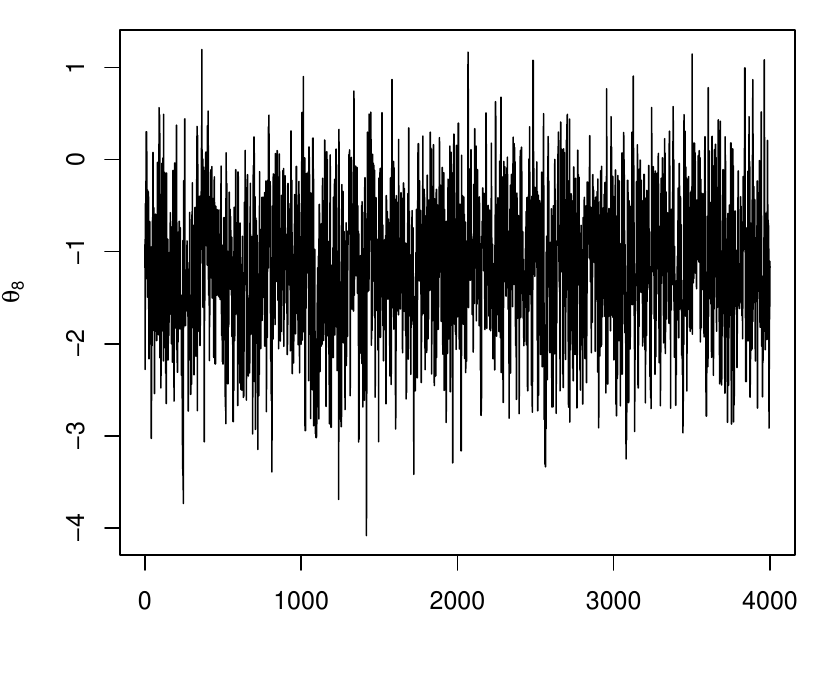} 
\caption{Trace plots for the posterior samples, where $\pmb{\theta}=(3, 1.5, 0, 0, 2, 0, 0, 0)$, errors follow the skew Student $t$ distribution, and $n=50$ in Simulation 1.}
\label{sfig:sim1theta}
\end{figure}
 
\begin{figure}[!ht]
\centering
\includegraphics[width=0.35\linewidth]{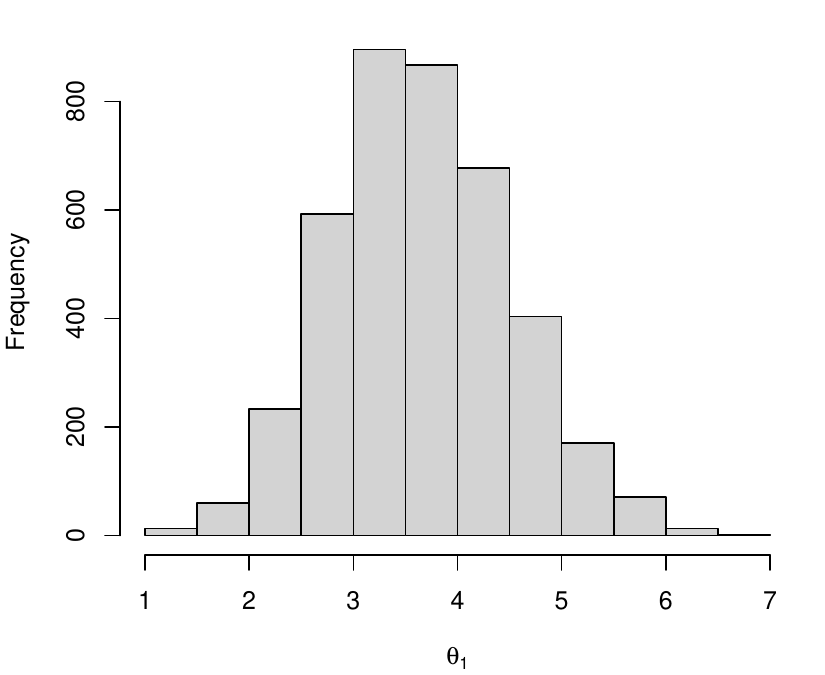} 
\includegraphics[width=0.35\linewidth]{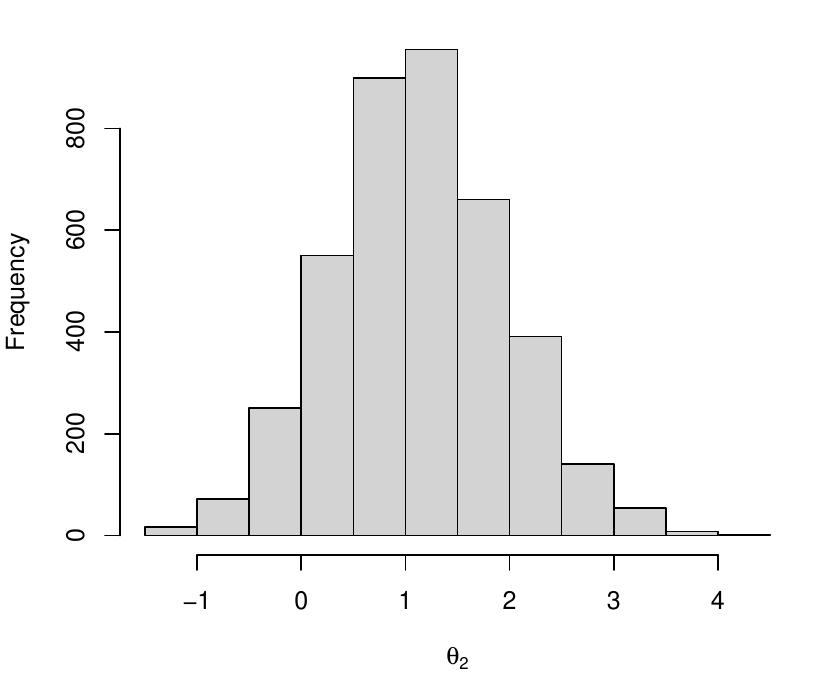} 
\includegraphics[width=0.35\linewidth]{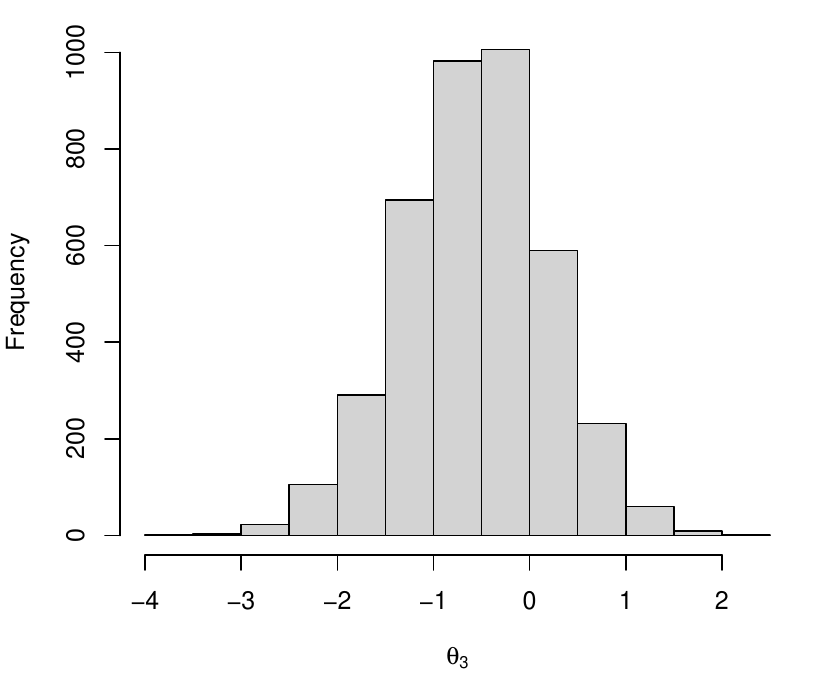} 
\includegraphics[width=0.35\linewidth]{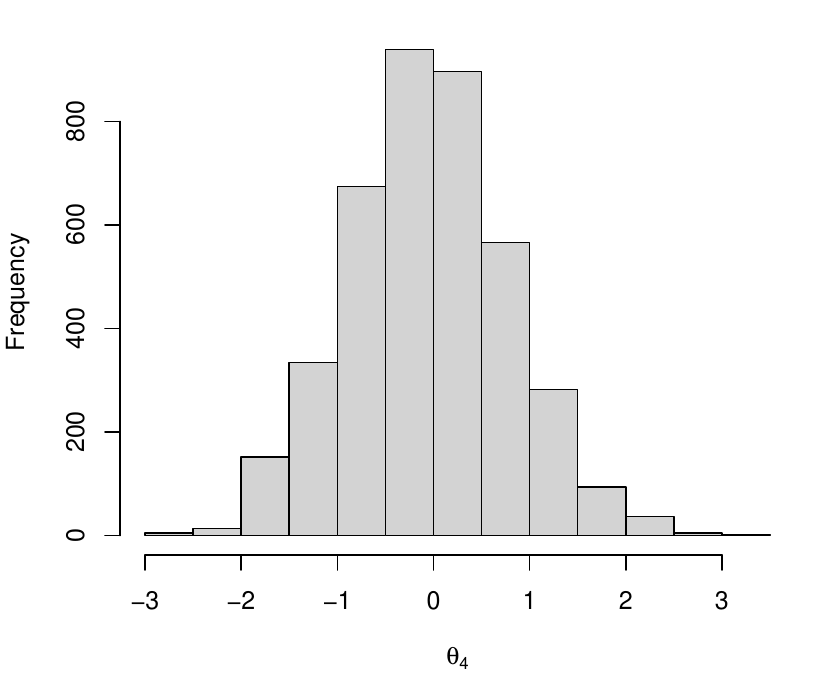} 
\includegraphics[width=0.35\linewidth]{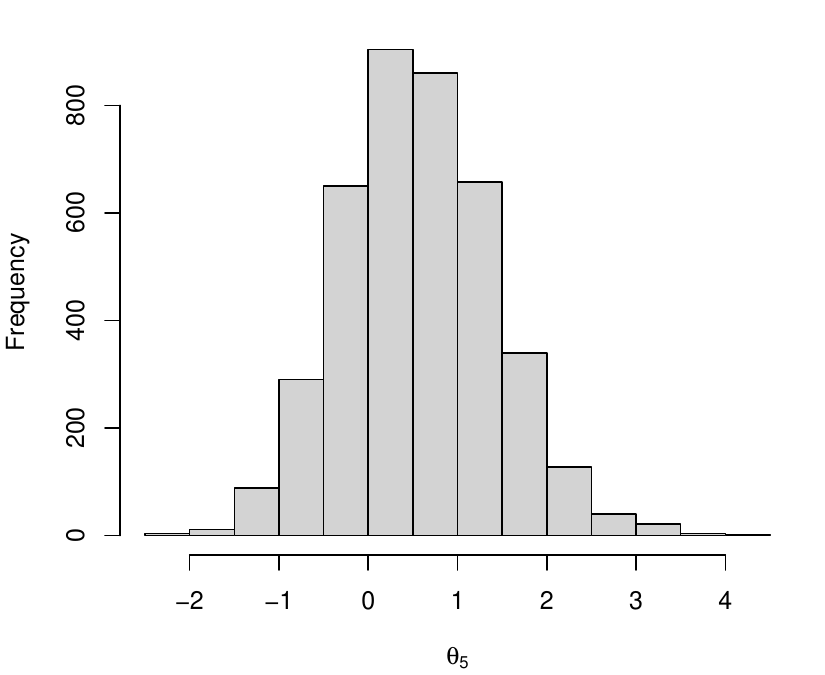} 
\includegraphics[width=0.35\linewidth]{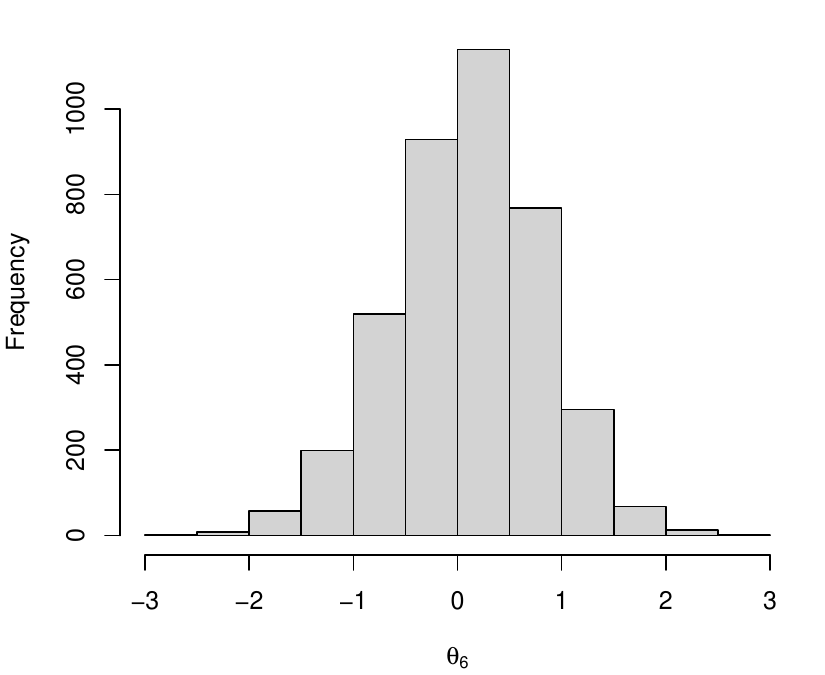} 
\includegraphics[width=0.35\linewidth]{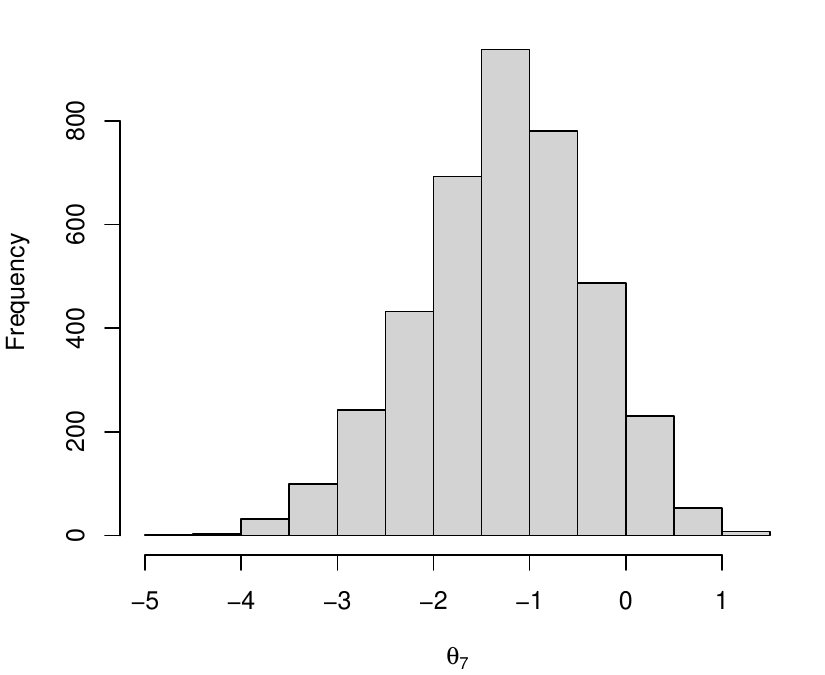} 
\includegraphics[width=0.35\linewidth]{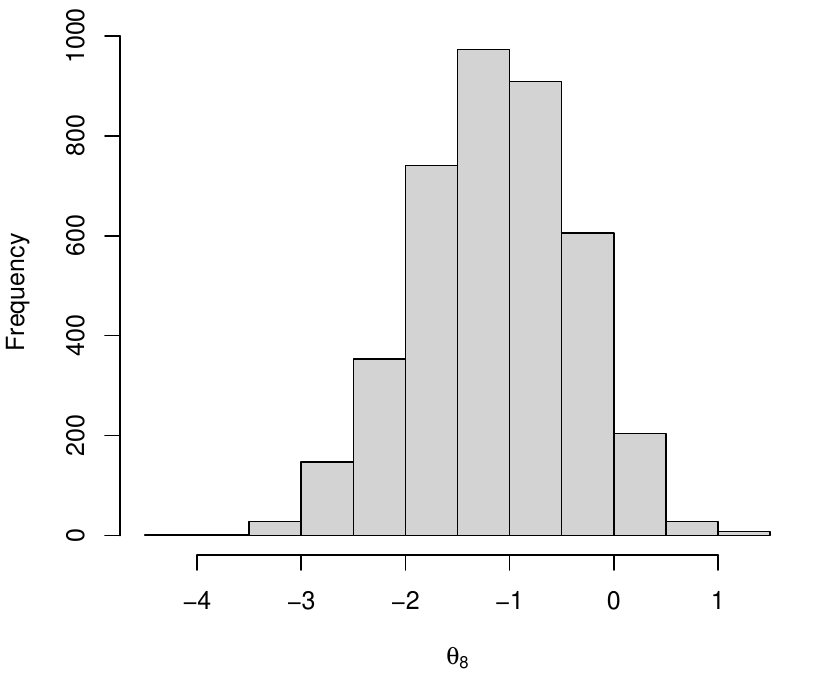} 
\caption{Histograms of posterior samples, where $\pmb{\theta}=(3, 1.5, 0, 0, 2, 0, 0, 0)$, errors follow the skew Student $t$ distribution, and $n=50$ in Simulation 1.}
\label{sfig:sim1hist}
\end{figure}

\begin{figure}[!ht]
\centering
\includegraphics[width=0.35\linewidth]{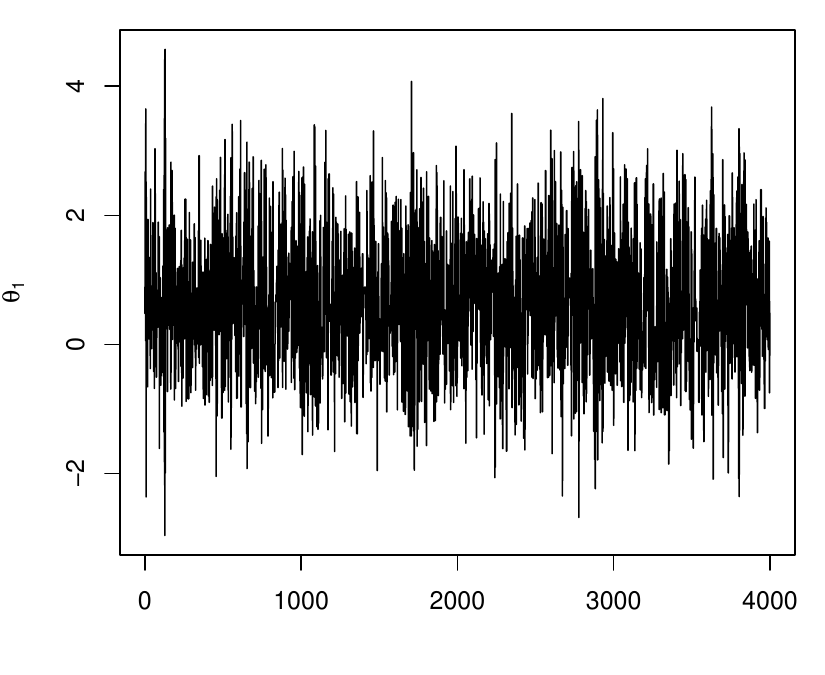} 
\includegraphics[width=0.35\linewidth]{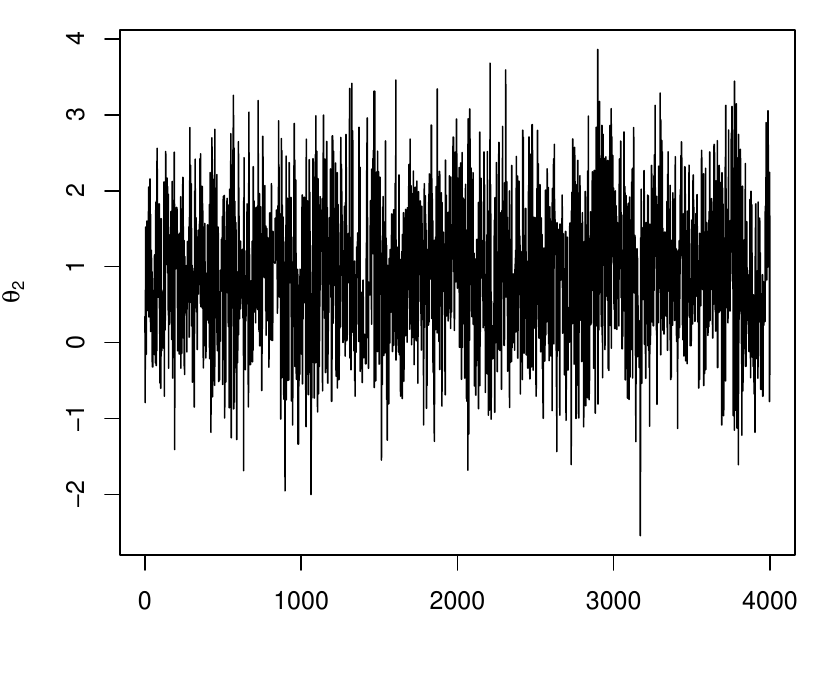} 
\includegraphics[width=0.35\linewidth]{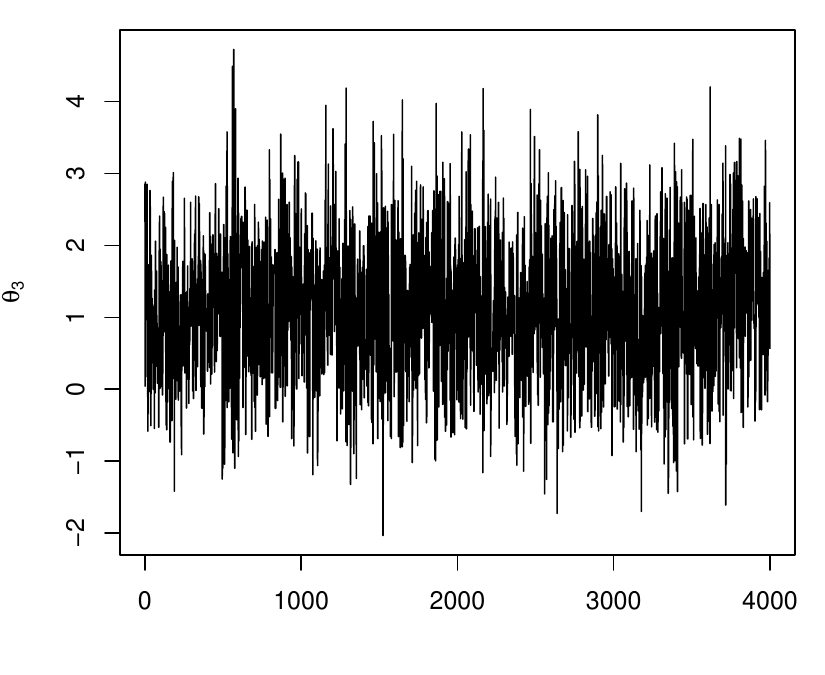} 
\includegraphics[width=0.35\linewidth]{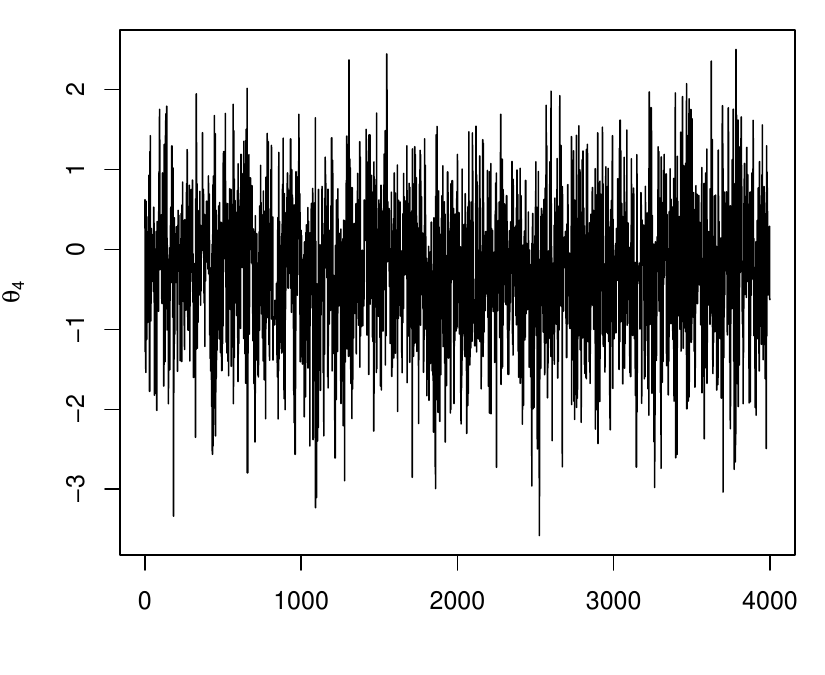} 
\includegraphics[width=0.35\linewidth]{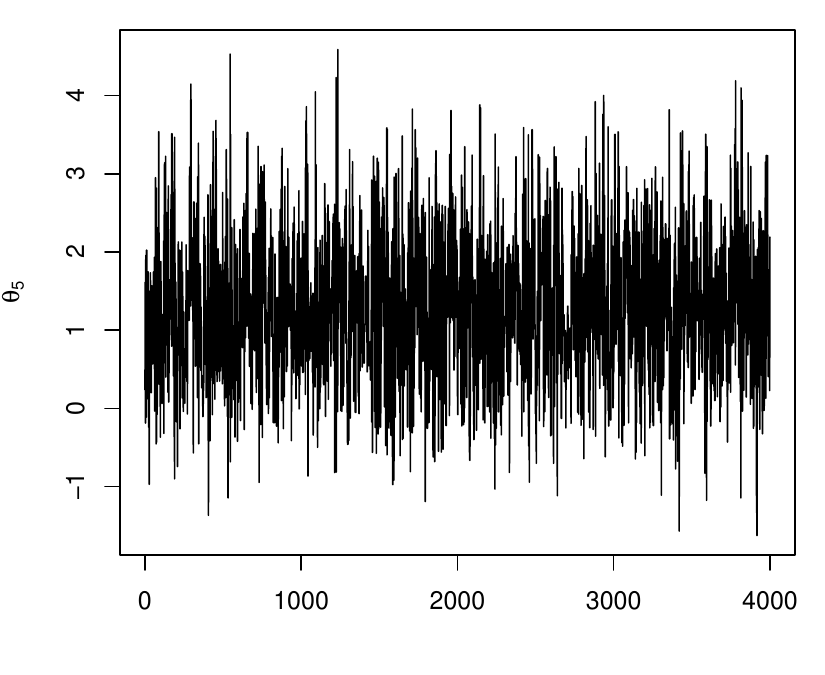} 
\includegraphics[width=0.35\linewidth]{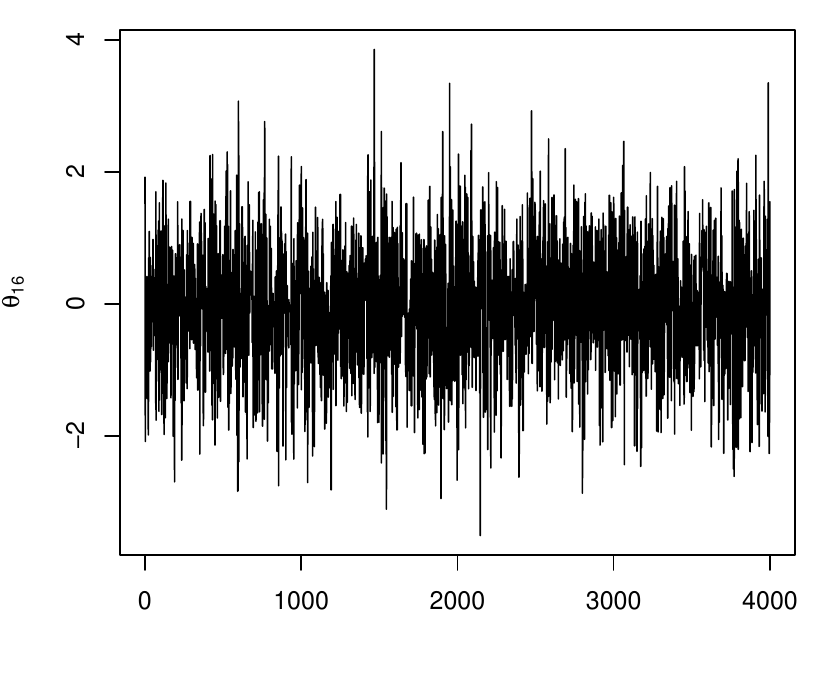} 
\includegraphics[width=0.35\linewidth]{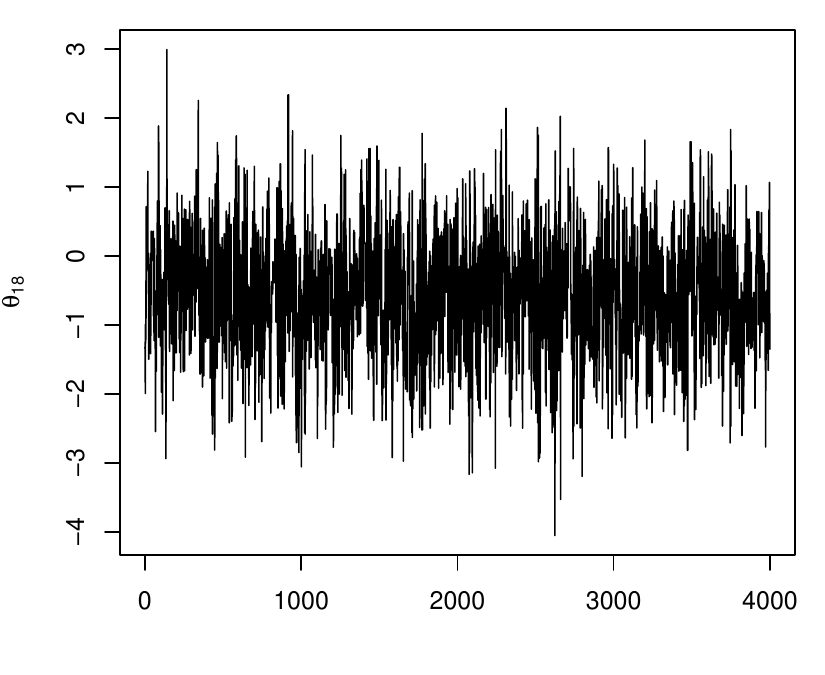} 
\includegraphics[width=0.35\linewidth]{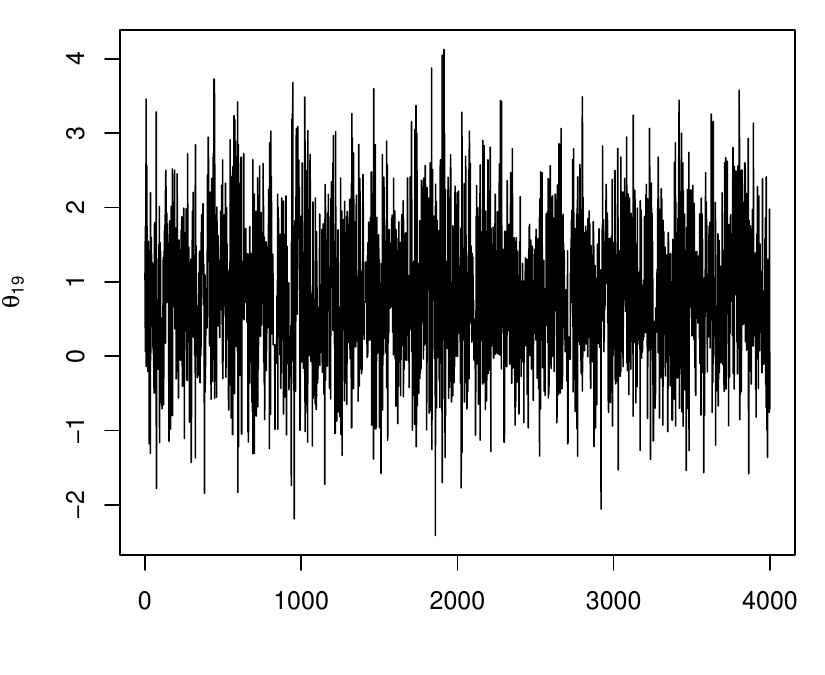} 
\caption{Trace plots for the posterior samples, where $\theta_1=\theta_2=\theta_3=\theta_4=3$,  $\theta_{16}=\theta_{17}=\theta_{18}=\theta_{19}=0)$, errors follow the skew Student $t$ distribution, and $n=100$ in Simulation 2.}
\label{sfig:sim2theta}
\end{figure}
 
\begin{figure}[!ht]
\centering
\includegraphics[width=0.35\linewidth]{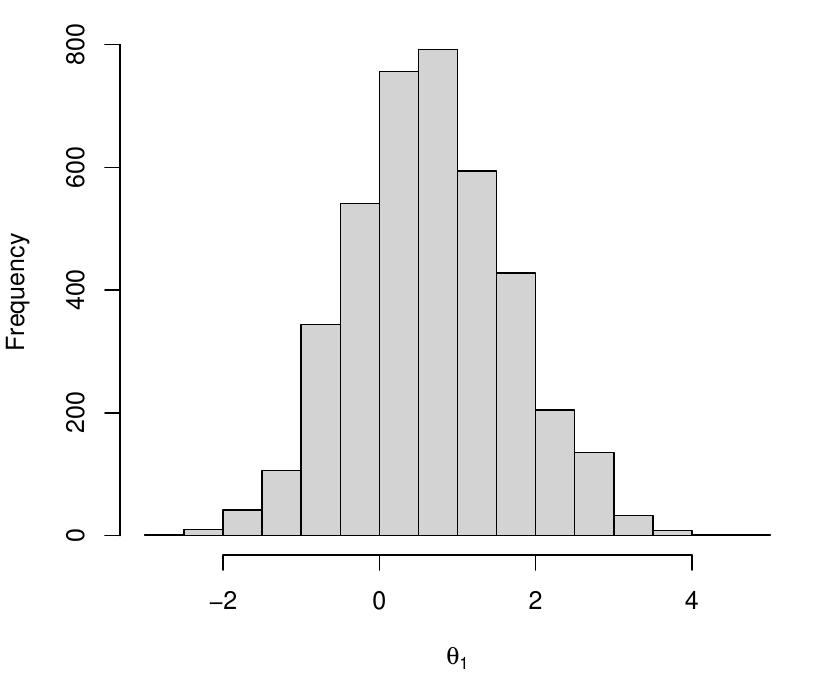} 
\includegraphics[width=0.35\linewidth]{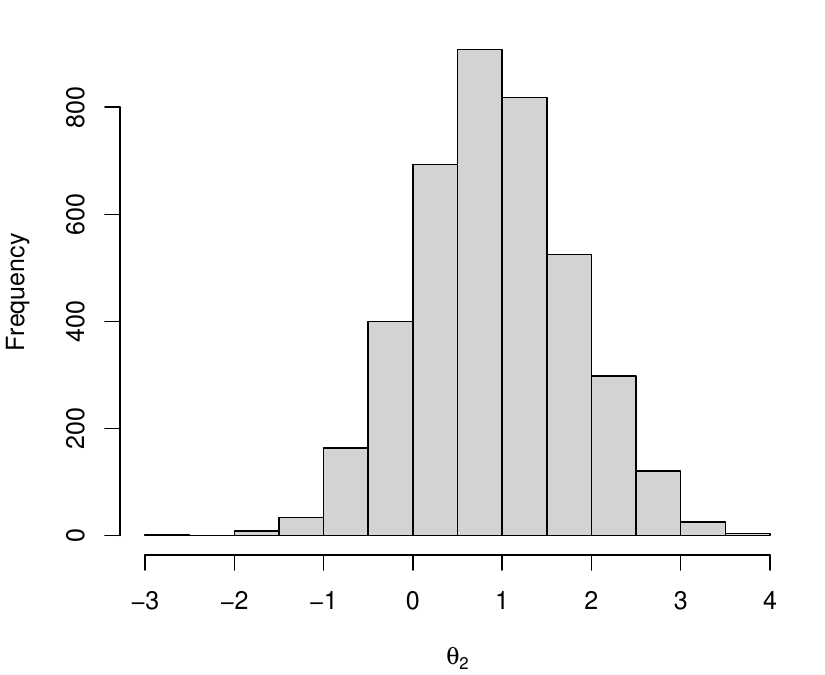} 
\includegraphics[width=0.35\linewidth]{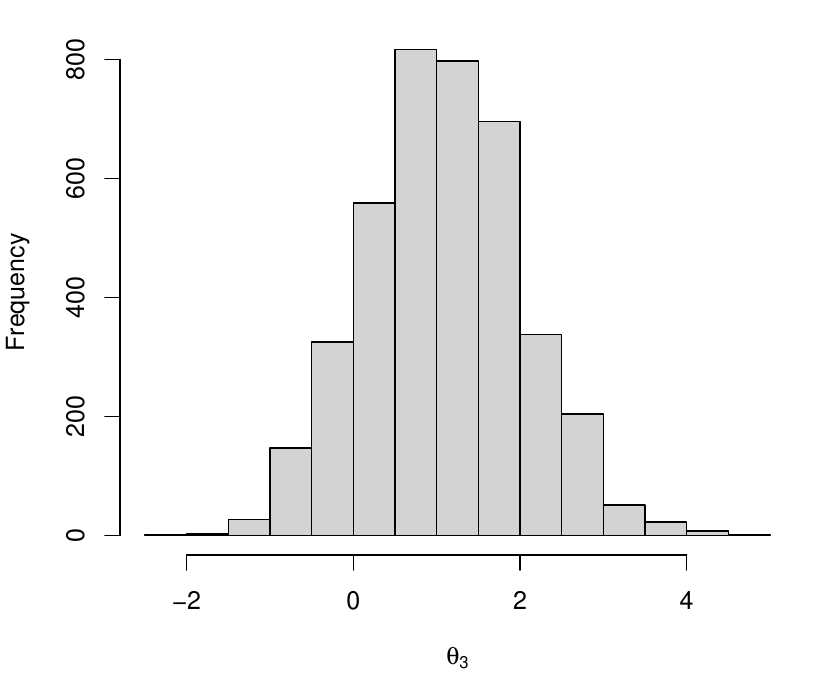} 
\includegraphics[width=0.35\linewidth]{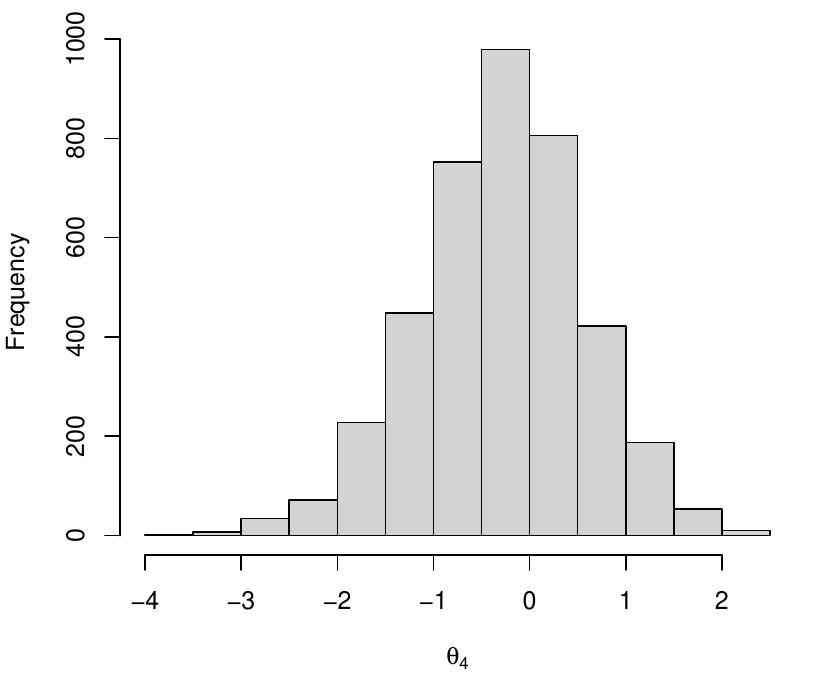} 
\includegraphics[width=0.35\linewidth]{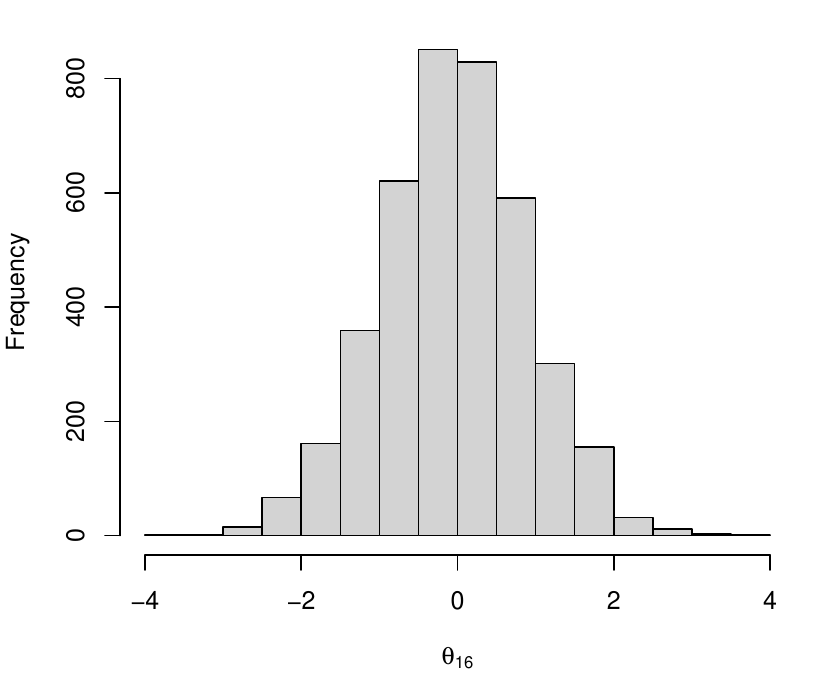} 
\includegraphics[width=0.35\linewidth]{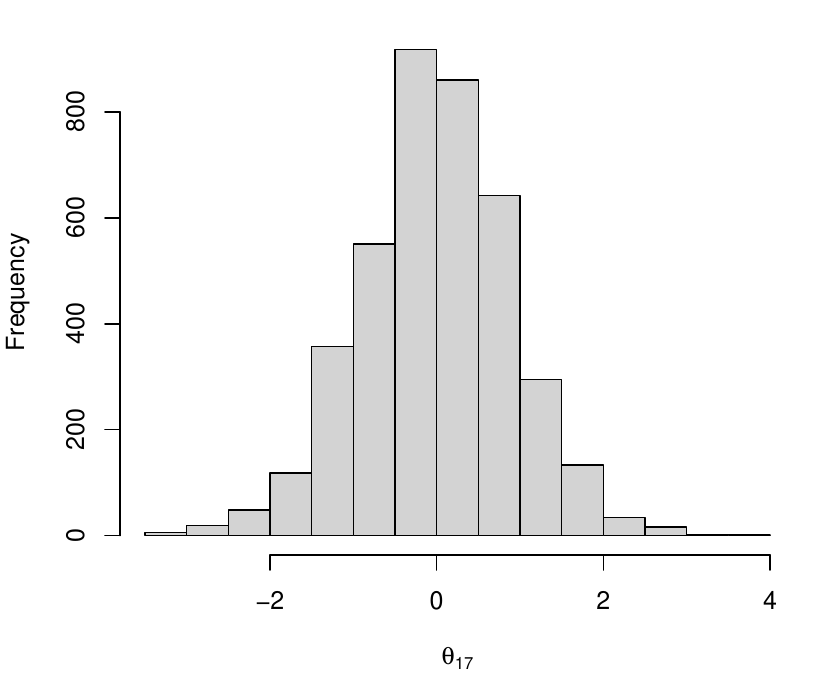} 
\includegraphics[width=0.35\linewidth]{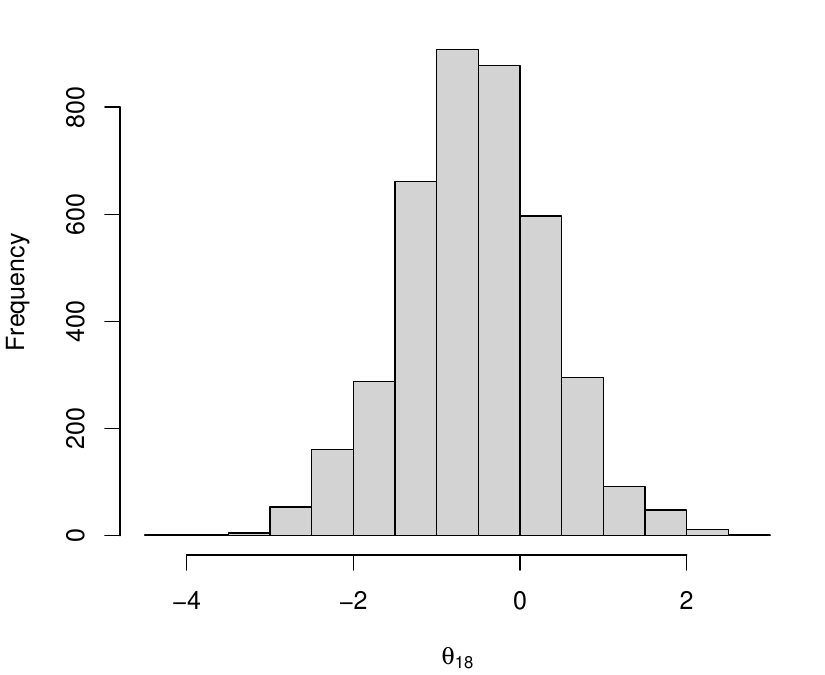} 
\includegraphics[width=0.35\linewidth]{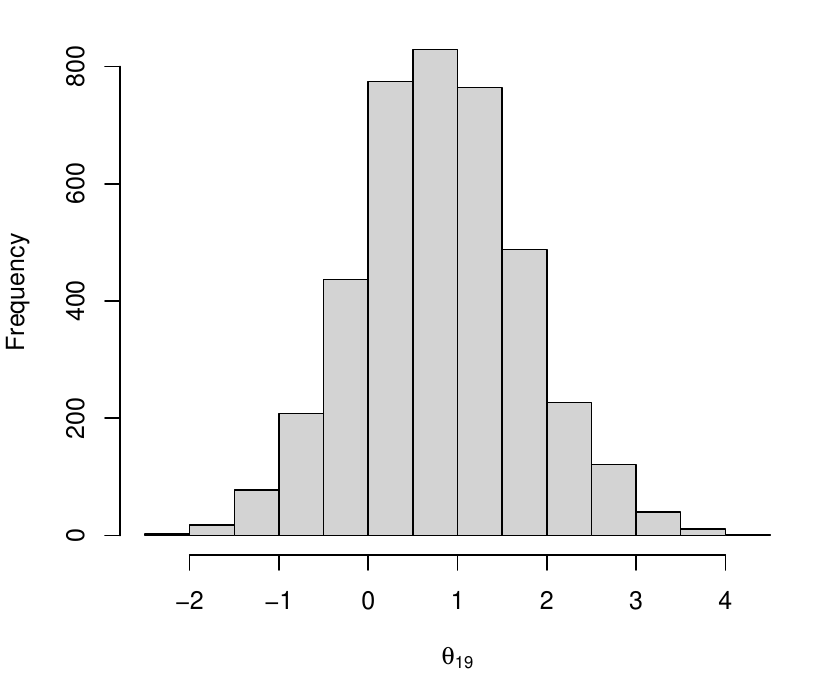} 
\caption{Histograms of posterior samples, where $\theta_1=\theta_2=\theta_3=\theta_4=3$,  $\theta_{16}=\theta_{17}=\theta_{18}=\theta_{19}=0)$, errors follow the skew Student $t$ distribution, and $n=100$ in Simulation 2.}
\label{sfig:sim2hist}
\end{figure}

%\newpage

%\section{Air pollution data analysis results}
%\label{appendix:b}

\begin{table}[!ht]
\tbl{Summary of variables of the air pollution data set.}
{
\begin{tabular}{|l|l|l|}
\hline
Variable type                  & Variable name & Description                                                 \\ \hline
\multirow{4}{*}{Weather}       & prec          & Mean annual precipitation (in)                              \\ \cline{2-3} 
                               & jant          & Mean January temperature (F)                                \\ \cline{2-3} 
                               & jult          & Mean July temperature (F)                                   \\ \cline{2-3} 
                               & humid         & Annual average relative humidity at 1 pm (\%)               \\ \hline
\multirow{8}{*}{Socioeconomic} & ovr95         & Percentage of population aged 65 or older in 1960           \\ \cline{2-3} 
                               & popn          & Population per household in 1960                            \\ \cline{2-3} 
                               & educ          & Median school years completed                               \\ \cline{2-3} 
                               & hous          & Percentage of housing units                                 \\ \cline{2-3} 
                               & dens          & Population per square mile in 1960                          \\ \cline{2-3} 
                               & nonw          & Percentage non-white population in 1960                     \\ \cline{2-3} 
                               & wwdrk         & Percentage employed in white collar occupations in 1960     \\ \cline{2-3} 
                               & poor          & Percent of families with income under 3,000 dollars in 1960 \\ \hline
\multirow{3}{*}{Pollution}     & hc            & Relative hydrocarbon pollution potential                    \\ \cline{2-3} 
                               & nox           & Relative oxides of nitrogen pollution potential             \\ \cline{2-3} 
                               & so            & Relative sulfur dioxide pollution potential                 \\ \hline
                               & mort          & Age-adjusted mortality rate per 100,000                     \\ \hline
\end{tabular}
}
\label{stab:air}
\end{table}

\begin{figure}[!ht]
\centering
\includegraphics[width=0.65\linewidth]{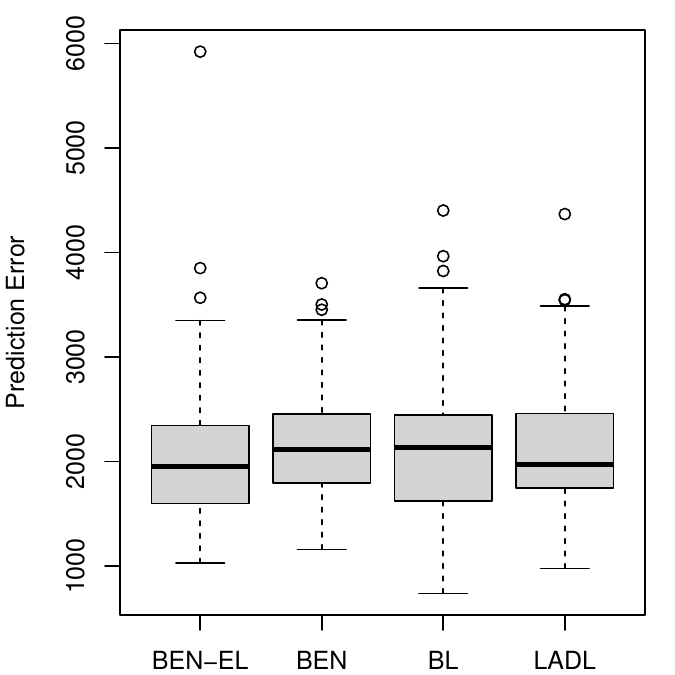} 
\caption{Boxplots of prediction errors for the air pollution data. BEN-EL has the smallest mean, median, Q1, and Q3. The prediction error of EN is not included because it differs from the other methods.}
\label{sfig:box}
\end{figure}

%\processdelayedfloats %%% See above for an explanation of why this command might be needed here.

\end{document}